\documentclass[10pt]{article}
\usepackage[normalem]{ulem}
\usepackage[utf8]{inputenc}
\usepackage[english]{babel}
\usepackage[toc,page]{appendix}

\usepackage{amsmath,amsfonts,amssymb,amsthm}
\usepackage{graphicx}
\usepackage{float}
\usepackage{hyperref}
\usepackage{subcaption}
\usepackage[format=plain,font=small,textfont=it]{caption}
\usepackage[dvipsnames]{xcolor}
\usepackage{mathrsfs}
\usepackage{physics}
\usepackage{empheq} 
\usepackage{dsfont} 
\usepackage{algorithm} 
\usepackage{enumitem}

\usepackage{listings}

\usepackage[noend]{algpseudocode}

\graphicspath{{pic/}}
\usepackage[colorinlistoftodos,prependcaption,textsize=footnotesize]{todonotes}

\newcommand{\sltc}{{$SL(2,\mathbb{C})$}}
\newcommand{\SUT}{{$SU(2)$}}

\newcommand{\new}[1]{{\color{black} #1} }

\newcommand{\Wthree}[6]{\left(\begin{array}{ccc} #1 & #2 & #3 \\ #4 & #5 & #6 \end{array}\right)}
\newcommand{\Wfour}[9]{\left(\begin{array}{cccc} #1 & #2 & #3 & #4 \\ #5 & #6 & #7 & #8 \end{array}\right)^{(#9)}}
\newcommand{\Wsix}[6]{\left \{ \begin{array}{ccc} #1 & #2 & #3 \\ #4 & #5 & #6 \end{array}\right \} }

\usepackage[usecourier=false,theme=jupyter,charsperline=115]{jlcode}

\begin{document}

\title{How-To compute EPRL spin foam amplitudes}

\author{\Large{Pietro Don\`a${}^{a}$\footnote{dona.pietro@gmail.com}, \ } \Large{Pietropaolo Frisoni${}^{b}$\footnote{pfrisoni@uwo.ca} \ }
\smallskip \\ 
\small{\textit{a Center for Space, Time and the Quantum, 13288 Marseille, France}}\\
\small{\textit{b Department of Physics and Astronomy, University of Western Ontario, London, ON N6A 5B7, Canada}}
}

\date{}

\maketitle

\begin{abstract}
\noindent Spin foam theory is a concrete framework for quantum gravity where numerical calculations of transition amplitudes are possible. 
Recently, the field became very active, but the entry barrier is steep, mainly because of its unusual language and notions scattered around the literature. This paper is a pedagogical guide to spin foam transition amplitude calculations. We show how to write an EPRL-FK transition amplitude, from the definition of the 2-complex to its numerical implementation using \texttt{sl2cfoam-next}. We guide the reader using an explicit example balancing mathematical rigor with a practical approach. We discuss the advantages and disadvantages of our strategy and provide a novel look at a recently proposed approximation scheme.
\end{abstract}


\section{Introduction}
\label{sec:intro}

Spin foam theory provides a background-independent, Lorentz covariant path integral for general relativity. Spin foams provide dynamics to Loop Quantum Gravity, defining transition amplitudes between spin network states. 
A triangulation discretizes the space-time manifold, and its 2-complex regularizes the partition function. 

The EPRL-FK model \cite{Engle:2007wy,Freidel:2007py} (we will refer to it as just EPRL for brevity) is the state-of-the-art spin foam model. There is a large consensus in the community \cite{Dona:2020tvv,Engle:2020ffj,Asante:2020qpa,Han:2021kll, Engle:2021xfs} that the classical continuum theory can be recovered with a double limit of finer discretization and vanishing $\hbar$.
This observation is supported by the emergence of Regge geometries and the Regge action in the asymptotics of the 4-simplex vertex amplitude for large quantum numbers \cite{article_Barrett_etal_2010_lorentzian_spinfoam_amplitude,Dona2020} and the recent study of many vertices transition amplitudes.

The amount of calculations possible within the models recently grew considerably. It was possible because of a paradigm shift in the field. It evolved from a theoretical framework to circumvent the difficulties in imposing the Hamiltonian constraint in the canonical approach \cite{Reisenberger:1996pu} into a concrete tool where numerical calculations of transition amplitudes are possible \cite{Dona2019, Dona2018, frisoni2021numerical, Sarno2018, frisoni2021studying,  Bahr:2017klw,Allen:2022unb}.

With increased interest in the field, its entry barrier also increased vastly. Getting into spin foam is very difficult for a student or a researcher from a different field. There are plenty of reviews \cite{Perez:2012wv,Ashtekar:2021kfp} and excellent books \cite{Rovelli2015} to study and learn the basic theory. On the other end of the spectrum, we have plenty of advanced frontline papers that explore the connection of spin foam with GR \cite{Dona:2020tvv, Spinfoam_Lefschetz_thimble, Han:2021kll} or possible phenomenological implications \cite{BW_part_1, article:Christodoulou_Rovelli_Speziale_Vilensky_2016_planck_star_tunneling_time, Gozzini_primordial, Cosm_project}. 

We noticed a hole in the literature. There are no papers that give you all the tools needed to complete a spin foam calculation, from its conception to the number. With this paper, we guide the reader through the calculation of an EPRL transition amplitude in a pedagogical manner. We use an explicit example to help them not feel disoriented dealing with abstract concepts. We hope that this paper can fill that hole and open spin foam calculation to a new generation of students and researchers. 

\new{To read this paper, advanced background knowledge on spin foam is not necessary. However, a basic understanding of the topic is helpful. We think of this work as a guide for making spin foam calculations. We refer to targeted reviews of the EPRL model \cite{Perez:2012wv,Rovelli2015,Ashtekar:2021kfp} for a comprehensive discussion of its definition, motivation, and physical significance. 

We start with a brief review of the construction of the spin foam theory and the definition of the EPRL model in Section~\ref{sec:theory}. In the rest of the paper, we show the reader how to compute a spin foam transition amplitude associated with a triangulation of the space-time manifold. We identify five necessary steps, each illustrated in a different section.
\begin{itemize}[leftmargin=17.5mm,labelsep=4.5mm]
\item[\textbf{Step~1.}] 
Draw the 2-complex.\\
In Section~\ref{sec:skeleton}, we describe how to build the 2-complex from the triangulation. It is crucial in regularizing the gravitational path integral and writing a finite transition amplitude.
\item[\textbf{Step~2.}] 
Write the EPRL spin foam amplitude.\\
In Section~\ref{sec:amplitude}, we give the prescription to write the transition amplitude associated with a 2-complex, and we introduce a very convenient graphical method to represent the amplitude. For the calculation, we resort to a divide-and-conquer strategy.
\item[\textbf{Step~3.}]  
Divide the EPRL transition amplitude into vertex contributions.\\
In Section~\ref{sec:manyvertices}, we show how to divide any transition amplitude into vertex amplitudes.
\item[\textbf{Step~4.}]  
Compute the EPRL vertex amplitudes.\\
In Section~\ref{sec:single_vertex}, we discuss the calculation of the vertex amplitude in terms of \SUT{} invariants and booster functions.
\item[\textbf{Step~5.}]  
Use \texttt{sl2cfoam-next} to compute a number.\\
We perform the numerical evaluation of the amplitude in Section~\ref{sec:numerical} using the numerical library \texttt{sl2cfoam-next} and discuss the necessary approximations. 
In this section, we also discuss and improve the extrapolation scheme discussed in~\cite{frisoni2021numerical} as a tentative to lift, at least part of, the approximation used to calculate the amplitude.
\end{itemize}
We complement our discussion with an explicit example. We compute the EPRL transition amplitude based on the triangulation $\Delta_4$. It was considered first in \cite{Asante:2021zzh} in Lorentzian the spin foams. It is 2-complex that at the same time not trivial (with more than one vertex), simple (with four vertices and some symmetry) but rich enough (with one bulk face) to require a certain degree of optimization to compute the associated amplitude. Moreover, in \cite{Asante:2021zzh} coherent boundary data corresponding to a Lorentzian geometry was provided, allowing semiclassical calculation with some ease that we leave to future work. }


\section{Overview of the EPRL model}
\label{sec:theory}
Spin foam theory is a promising approach to quantize gravity. The goal is to define a path integral for general relativity in a non-perturbative and background-independent way. The spin foam partition function assigns transition amplitudes between spin network states, a basis of the Loop Quantum Gravity kinematical Hilbert space. For this reason, spin foam theory gives a dynamic to Loop Quantum Gravity, and it is often referred to as Covariant Loop Quantum Gravity \cite{Rovelli2015}.

In the Plebanski formulation of general relativity \cite{Plebanski:1977zz}, we formulate gravity as a topological BF theory with constraints \cite{Baez:1999sr}. The variables of the BF theory are a 2-form $B$ conjugated to a connection $\omega$ (with curvature $F$)\footnote{This theory has no degrees of freedom: all the solutions of the equations of motion are gauge equivalent to the trivial one $d_\omega B = 0$ and $F(\omega) = 0$. The name derives from the name of the variables used and the simple form of the action $\int_\mathcal{M} B \wedge F(\omega)$.}. 

General relativity is not topological, the \emph{simplictiy} constraints reduce the B-field in BF theory to a $\gamma$-simple 2-form $B =  \star e \wedge e  +\tfrac{1}{\gamma} e \wedge e$ , reducing the action to the familiar Holst action \cite{Holst:1995pc}.

The path integral of spin foam theory is regularized on a triangulation, more precisely its 2-complex, to truncate the degrees of freedom. We discretize and quantize the topological theory first. The $B$-fields are assigned to faces of the 2-complex, triangles, and encode their geometry.
The connection is regularized by considering only its holonomy $g$ responsible for the parallel transport along the (half-)edges of the 2-complex (from one tetrahedra to another). The topological theory partition function consists of a collection of delta functions imposing flatness of each face of the 2-complex.

The partition function of the EPRL model is derived enforcing the simplicity constraints at the quantum level to reduce the topological theory to gravity. On a simplicial triangulation, we have a \emph{linear} version of the simplicity constraints: we require the proportionality between the boost and rotation generators of \sltc{} $\vec{K} = \gamma \vec{L}$ at the boundary of any 4-simplex (vertex of the 2-complex). The generalization to arbitrary tessellation is possible \cite{Kaminski:2009fm} but requires complications beyond this work's scope. Therefore, we limit ourselves to 4-simplices.

The key ingredient of the EPRL model is the $Y_\gamma$ map. It embeds the spin $j$ $SU(2)$ representation into the lowest spin sector of the unitary irreducible representations in the principal series of \sltc{} labeled by $\rho,k = \gamma j,j$.

We expand the BF theory partition function in terms of matrix elements of the holonomies in irreducible rerpesentations of \sltc{} $D^{\rho, k}_{jmln}(g)$. See Appendix~\ref{app:sltc} and references therein for more details. The EPRL model prescription enforces the $Y_\gamma$ map at every vertex of the 2-complex restricts the irreducible representations to $\gamma$-simple ones $D^{\gamma j, j}_{jmjn}(g)$ \cite{Engle:2007wy}. 

If the 2-complex has a boundary, the spin foam partition function maps states from the Loop quantum Gravity kinematical Hilbert space (identified with the boundary space of the spin foam with the $Y_\gamma$ map) into the complex numbers (quantum transition amplitudes between these states). 

The EPRL spin foam partition function is given as a state sum over $SU(2)$ spins $j_f$ on the faces and intertwiners $i_e$ on the edges of the 2-complex: 
\begin{equation}
\label{eq:partitionf}
Z_{\Delta} = \sum_{j_f, i_e}  \prod_f A_f(j_f) \prod_e A_e(i_e) \prod_v A_v \left(j_f, \  i_e\right) \ ,
\end{equation}
defined in terms of the face amplitude $A_f$, and the edge amplitude $A_e$ and the vertex amplitude $A_v$. Requiring the correct convolution property of the path integral at fixed boundary, the form of the face amplitude $A_f(j_f) = 2 j_f +1$ and the edge amplitude $A_e(i_e)=2i_e+1$ are fixed \cite{Bianchi:2010fj}. 

We will not give an explicit form of the amplitude for an arbitrary 2-complex. They can be found in many references \cite{Engle:2007wy, Perez:2012wv, Rovelli2015} if the reader is interested. Instead, we opt for a constructive approach. In Section~\ref{sec:amplitude}, we guide the reader through a set of rules to write a general EPRL transition amplitude. In Section~\ref{sec:manyvertices}, we divide the transition amplitude in vertex amplitudes. In Section~\ref{sec:single_vertex}, we discuss the explicit form of the vertex amplitude and its form best suited for numerical calculations.

\section{How-To draw the 2-complex}
\label{sec:skeleton}
The spin foam partition function is regularized on the 2-complex of a triangulation of the space-time manifold. Given a triangulation, we can build its 2-complex associating a \emph{vertex} to each 4-simplex.

\smallskip

Each 4-simplex shares a tetrahedron with an adjacent 4-simplex. We associate to each tetrahedron of the triangulation an \emph{edge} of the 2-complex. An edge connects two adjacent vertices. Each triangle in a 4-simplex is shared by two tetrahedra, which are generally shared with other 4-simplices. In the whole triangulation, a triangle can be shared by any number of tetrahedra and 4-simplices.

\smallskip

We associate to each triangle of the triangulation a \emph{face} of the 2-complex. A face can contain any number of vertices and all the edges connecting them. We also assign an \emph{orientation} to the faces of the 2-complex. This choice is needed for a well-defined notion of parallel transport (to identify the source and target of the holonomy uniquely). 

\smallskip

Since two vertices share each edge for each of them, we can introduce two \emph{half-edges}, one associated with each vertex. We can still picture them as dual to the tetrahedron but ``seen'' in the 4-simplex it belongs to. In each vertex in a given face there are two half-edges. This is sometimes referred to as a \emph{wedge}. The orientation of the face allows us to identify one half-edge as the source tetrahedron (reference frame) and the other as the target tetrahedron (reference frame) of parallel transport along the face from the first one to the last one. If a boundary is present, the edges intersected by the boundary are severed in half, leaving only one half-edge in the skeleton.

\smallskip

We summarize the nomenclature introduced in this section in the following:
\smallskip
\begin{center}
    \sffamily
    \begin{tabular}{ccccc}
        &  \Large Triangulation &  & \Large 2-complex & \\ \hline\hline
        & & \\
         \raisebox{-0.5\height}{\includegraphics[width=1.5cm]{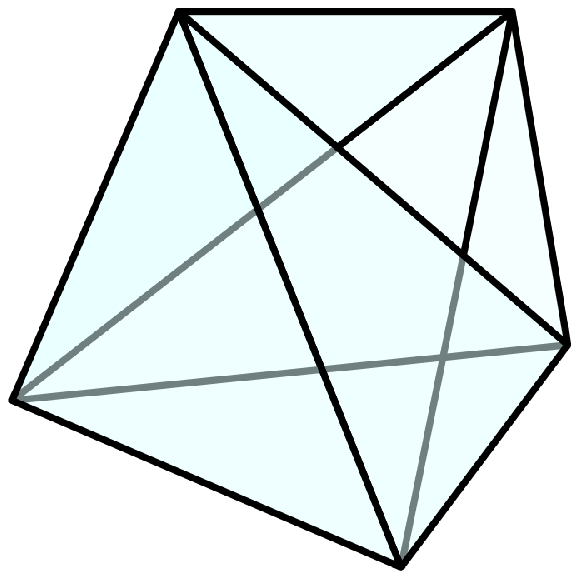}}    &
        4-simplex  & 
        $\rightleftarrows$ & 
        Vertex  & 
         \raisebox{-0.5\height}{\includegraphics[width=0.35cm]{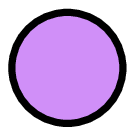}} \\
         \raisebox{-0.5\height}{\includegraphics[width=1.5cm]{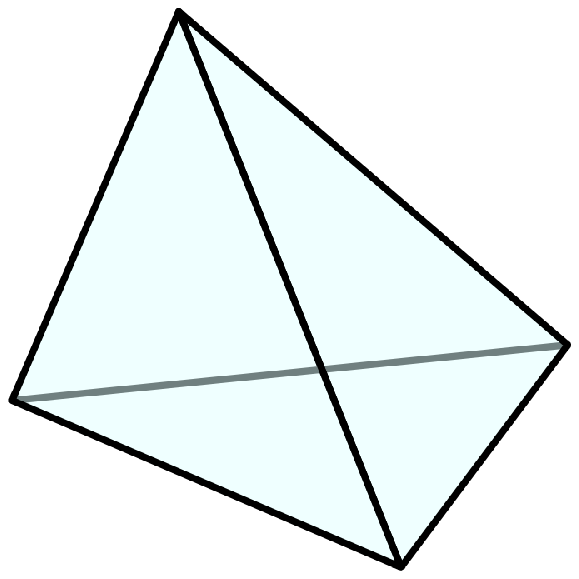}} & 
        Tetrahedron  & 
        $\rightleftarrows$ & 
        Edge & 
         \raisebox{-0.5\height}{\includegraphics[width=1.5cm]{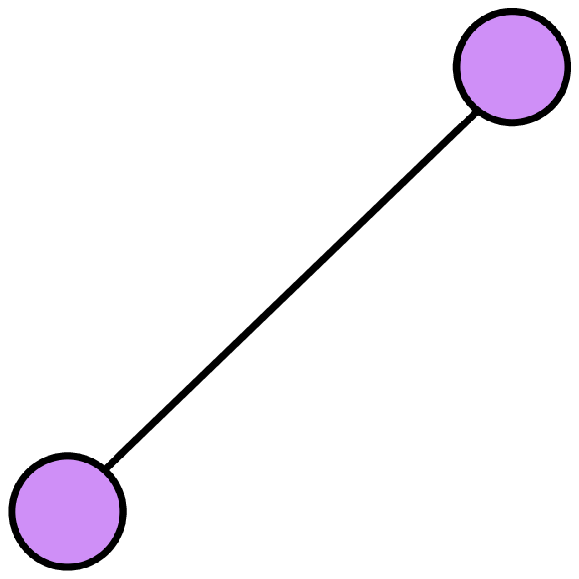}}    \\
         \raisebox{-0.5\height}{\includegraphics[width=1.5cm]{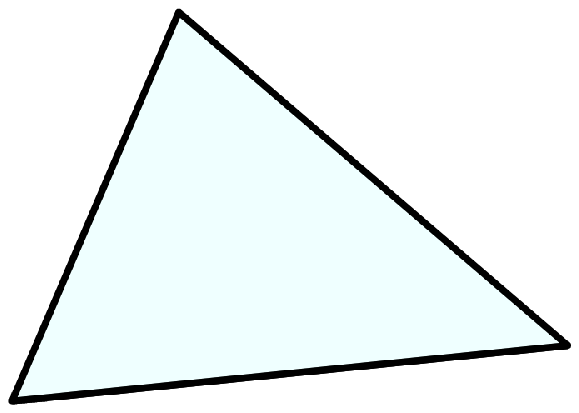}} &
        Triangle & 
        $\rightleftarrows$ & 
        Face & 
         \raisebox{-0.5\height}{\includegraphics[width=1.5cm]{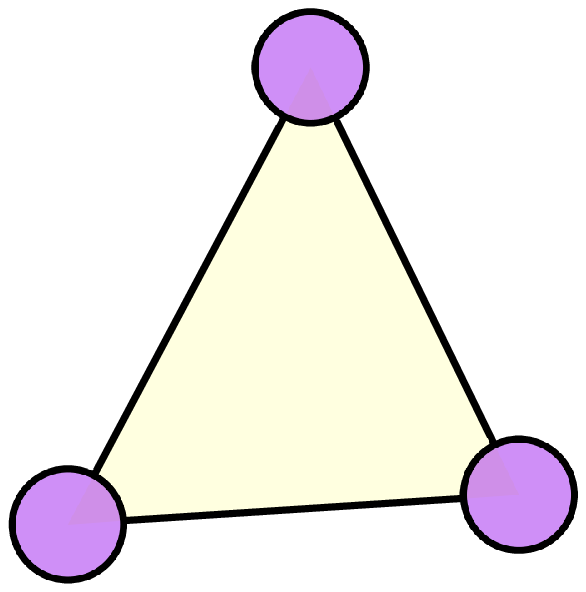}}    \\
         \raisebox{-0.5\height}{\includegraphics[width=1.5cm]{2skeleton/tetrahedron.eps}} &  \parbox[t]{3cm}{\centering Tetrahedron within 4-simplex} & 
        $\rightleftarrows$ &
        Half-edge    &  \raisebox{-0.5\height}{\includegraphics[width=0.85cm]{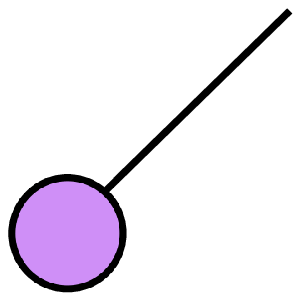}}    \\
        \raisebox{-0.7\height}{\includegraphics[width=1.5cm]{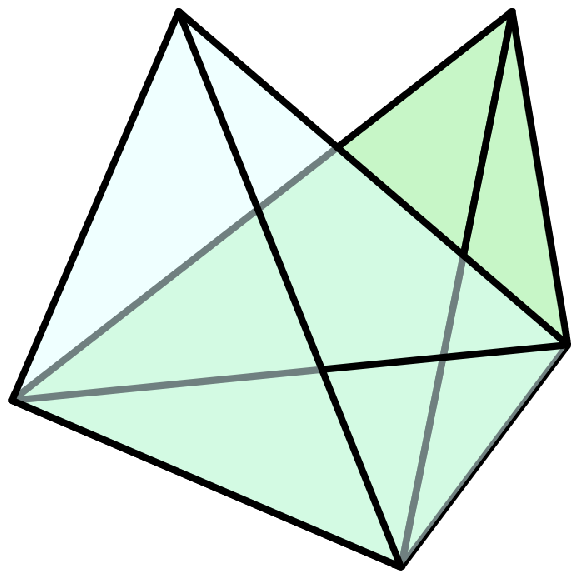}} &
        \parbox[t]{3cm}{\centering Oriented couple of tetrahedra in the same simplex} & 
        $\rightleftarrows$ & 
        Wedge & 
         \raisebox{-0.5\height}{\includegraphics[width=0.9cm]{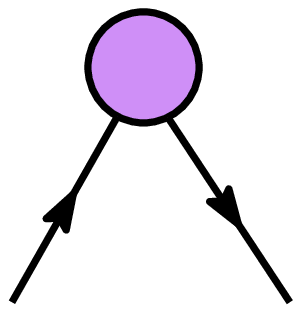}}    \\
        & & \\
        \hline\hline
    \end{tabular}
\end{center}

\begin{figure}[H]
  \centering
    \raisebox{-0.5\height}{\includegraphics[scale=0.56]{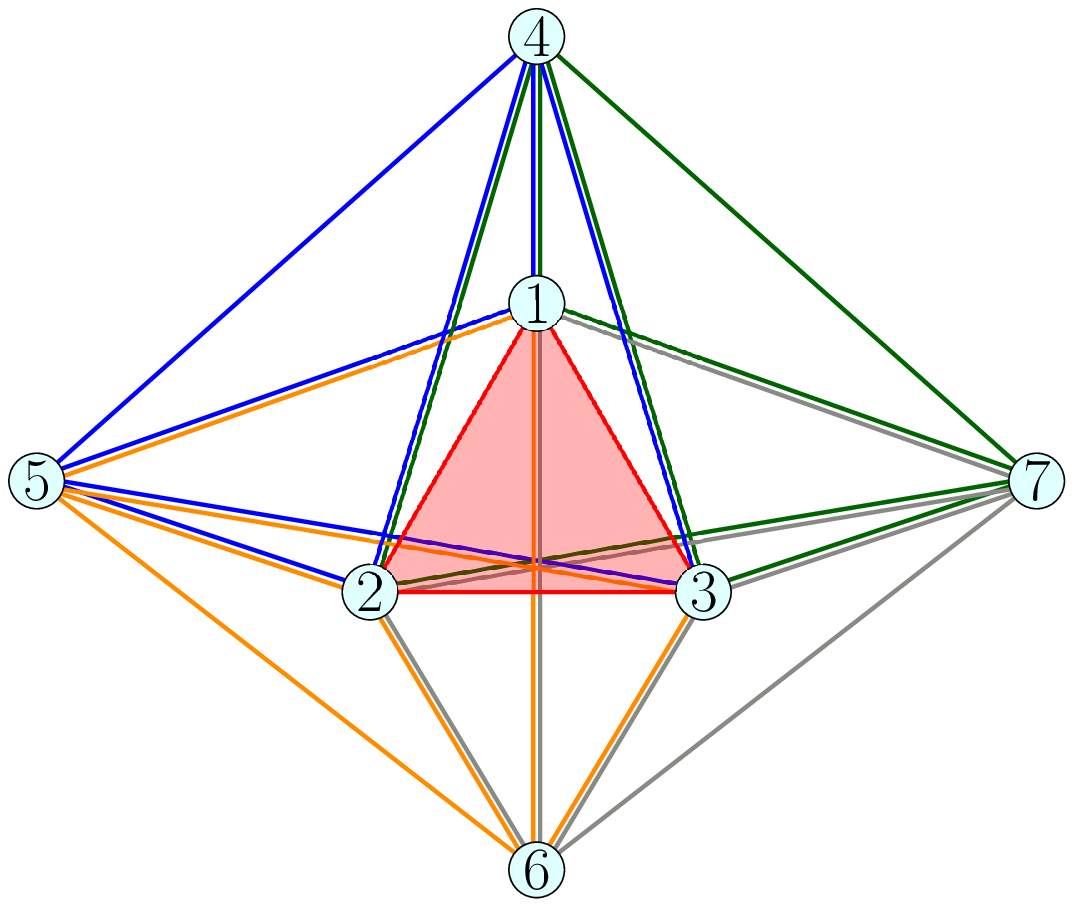}}
      \hspace{2em}
    \raisebox{-0.5\height}{\includegraphics[scale=0.56]{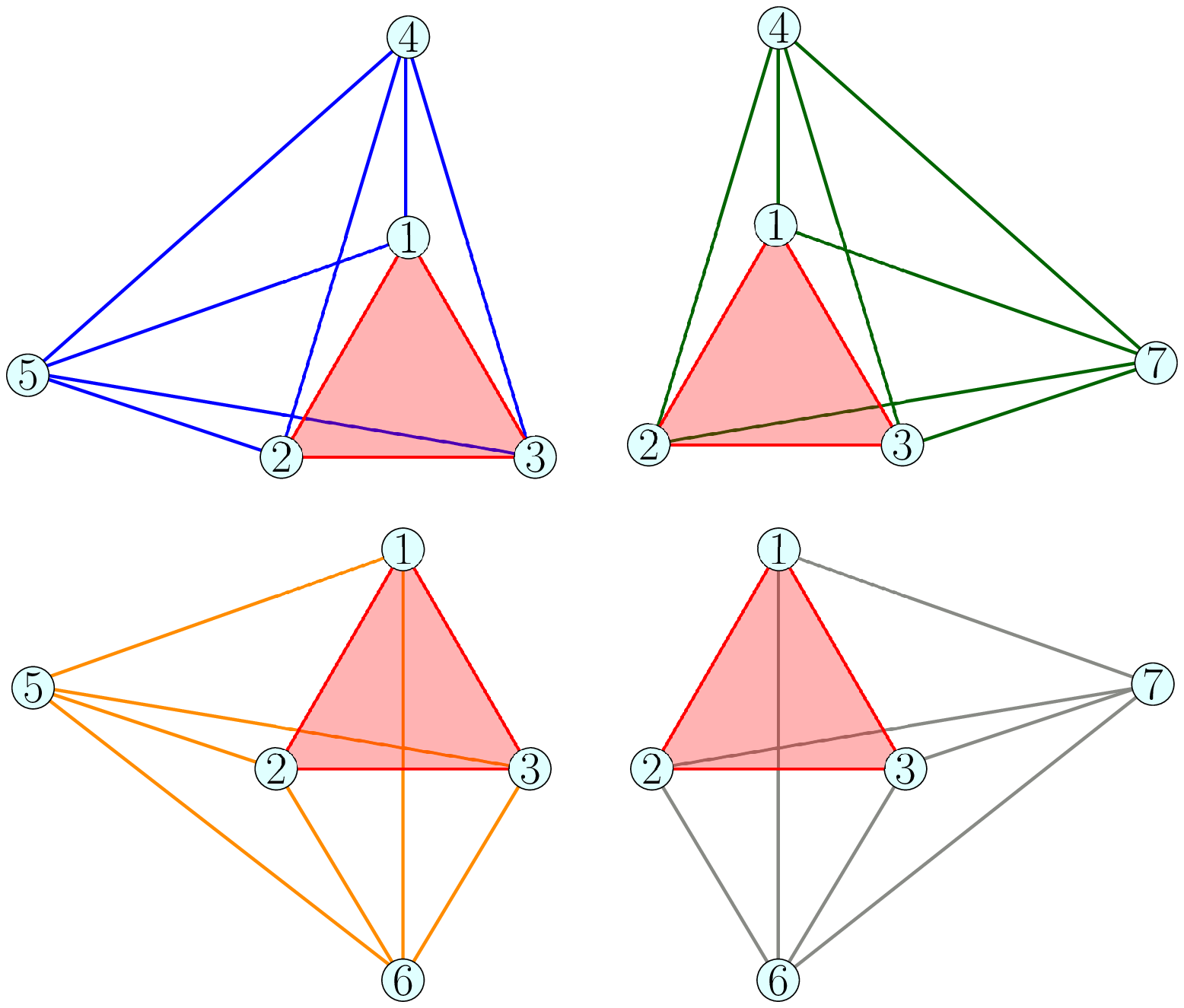}}
  \caption{The $\Delta_4$ triangulation. The numbered circles correspond to points, while lines correspond to segments. Each color corresponds to a different 4-simplex. The bulk triangle $123$ is highlighted in red. In the right panel, the 4-simplices are shown separately.}
  \label{fig:geometry}
\end{figure}

\new{\subsection{An example: the $\Delta_4$ triangulation}}
The triangulation is formed by four 4-simplices, all sharing a triangle. The triangulation has seven points, nineteen segments, twenty-five triangles, sixteen tetrahedra (twelve in the boundary and four in the bulk), and four 4-simplices. We label the points with numbers from 1 to 7, segments with couples of different numbers (points), triangles with triples of distinct numbers (the shared triangle is $123$), tetrahedra with a quadruple of distinct numbers, and 4-simplices with five distinct numbers. See Figure~\ref{fig:geometry} for a pictorial representation of the triangulation.

\begin{figure}
  \centering
  \includegraphics[scale=0.8]{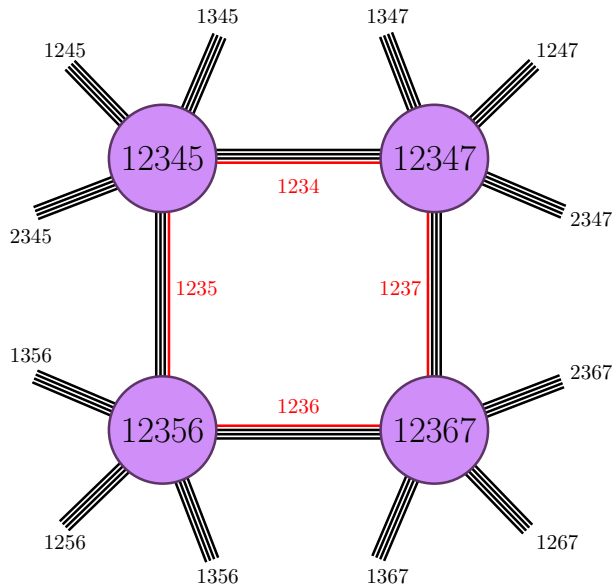}
  \caption{The 2-complex of the $\Delta_4$ triangulation. We named the vertices and the tetrahedra explicitly. We avoided naming the faces explicitly not to clutter the figure. Three numbers label the faces. We find a face's name looking for the numbers in common to all the edges it belongs to. For example, the tetrahedra $1234$, $1235$, $1236$, and $1237$ all share the face $123$.}
  \label{fig:2skeleton}
\end{figure}

The 2-complex of the $\Delta_4$ triangulation has four vertices associated with a 4-simplex. It has four internal edges, each associated with a tetrahedron shared among two 4-simplices. There are also three external edges for each vertex. Each edge belongs to 4 faces, and each face is associated with a triangle of the $\Delta_4$ triangulation. All triangles but one belong to the boundary of the triangulation. Therefore all faces but one of the 2-complex are boundary faces. The bulk face is associated with the triangle shared by all the 4-simplices. Thus it is crossing all four vertices. We label the 2-complex in the same way of the triangulation, see Figure~\ref{fig:2skeleton} for a representation. 

\section{How-To write the EPRL spin foam amplitude}
\label{sec:amplitude}

For each wedge we write a $\gamma$-simple unitary irreducible representation in the principal series of \sltc{} (see Appendix~\ref{app:sltc} and reviews \cite{Pierre_notes_LQG} and references therein for more mathematical details).
\begin{equation}
\label{eq:ymap}
D^{\gamma j, j}_{jm, jn} (g_w) \ , 
\end{equation}
where $j\in \mathbb{N}/2$ is a spin, $\gamma$ is the Immirzi parameter coming from the simplicity constraints, $m,n$ are magnetic indices $m,n=-j, -j+1 \ldots, j-1, j$, and $g_w \in SL(2,\mathbb{C})$ is a group element associated to the wedge. This restriction results from the weak quantum implementation of the simplicity constraints in the EPRL spin foam model. The $Y_\gamma$ map is responsible of this implementation and embeds the spin $j$ $SU(2)$ representation in \sltc{} as in \eqref{eq:ymap}.
The group element $g_w$ represents the holonomy responsible for the parallel transport along the wedge from the reference system of the source tetrahedron to the reference system of the target tetrahedron. We conventionally associate the row of the representation matrix, the couple $(j,m)$ in \eqref{eq:ymap}, to the target and the column, the couple $(j,n)$ in \eqref{eq:ymap}, to the source. In this way, the \sltc{} $\gamma$-simple representation matrices inherit the orientation of the 2-complex.

Instead of a group element for each wedge, we prefer to use a group element for each half-edge. We replace $g_w \to g_{t}^{-1} g_s $ where $s$ and $t$ are the source and target half-edges. 
This choice of fundamental variables guarantees that the parallel transport on a closed path in a vertex is trivial. In other words, the product of all the holonomies on the same closed path is the identity
\footnote{
Explicitly, if $w_1$, $w_2$ and $w_3$ are three wedges of the same vertex we have
\begin{align}
    g_{w_3}g_{w_2}g_{w_1}=&g_{e_1}^{-1} g_{e_3} g_{e_3}^{-1} g_{e_2} g_{e_2}^{-1} g_{e_1} = \mathds{1} \ , 
\end{align}
where we have assumed that the wedges are oriented such that the target of $w_1$ is the source of $w_2$ and so on. If the orientation of one of the wedges $w$ is the opposite we replace $g_w$ with its inverse.
}, or the holonomy is flat within a single vertex.

We set the spin $j$ on each edge to be the same and contract the magnetic indices $m$, $n$. At the end of this procedure, the only non-contracted magnetic indices are on boundary half-edges. We prescribe them as part of the boundary data. A common choice to describe boundary data is to contract these magnetic indices with intertwiners in the recoupling basis or with coherent intertwiners if we are interested in representing some semi-classical geometry.

We sum over all the possible spins $j_f$ associated to each closed face and we weight the contribution of the face with the dimensional factor $(2j_f+1)$ \cite{Bianchi:2010fj}
\begin{equation}
    \sum_{j_f} (2j_f+1) \sum_{m_w,n_w} \left(\prod_{w\subset f} D^{\gamma j_f, j_f}_{j_f m_w j_f n_w}(g_{w})\right) \ ,
\end{equation}
where the product is on all the wedges belonging to the face.
On non closed-faces we assign the spin as part of boundary data.  

We integrate over the group element associated to each half edge using the Haar measure of \sltc{}. For each vertex one integration is redundant and we remove it to regularize the amplitude as prescribed in \cite{Engle:2008ev}.
\new{\subsection{Graphical notation}}
Writing all the constituent of an EPRL spin foam amplitude can quickly get out of hand. To help us be precise and clear, we rely on a graphical notation. We introduce the various elements as we need them. We represent a unitary irreducible representation in the principal series of \sltc as an oriented line. The row labels correspond to the start of the line, and the column labels to the end of the line. We indicate the argument group element in a box and decorate the line with the needed representation labels
\begin{align}
\label{eq:irrepsl2c}
 D^{\rho, k}_{jm ln}(g)  \hspace{1em} & = \hspace{1em} \raisebox{-0.45\height}{\includegraphics[scale=1]{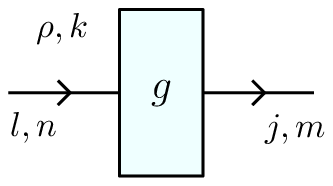}} \ .    
\end{align}
We contract two representations summing over all the magnetic indices (both $j$ and $m$ are magnetic numbers from the perspective of the infinite-dimensional irreducible representations of \sltc) by connecting the two lines. For example, in graphical notation, the \sltc{} representation property reads
\begin{align}
\label{eq:irrepsl2ccomp}
 D^{\rho,k}_{jm ln}(g_2 g_1) = \sum_{\substack{i\geq k\\ |p|\leq i }} D^{\rho,k}_{jm ip}(g_2) D^{\rho,k}_{ip ln}(g_1)  \hspace{1em} & = \hspace{1em} \raisebox{-0.45\height}{\includegraphics[scale=1]{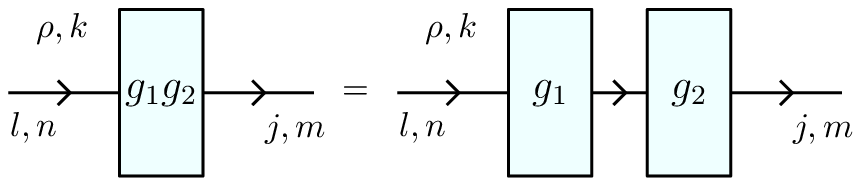}} \ .    
\end{align}
We denote the implementation of the $Y_\gamma$ map \eqref{eq:ymap} with a blue thick line that cuts across the representation line:
\begin{align}
\label{eq:graphical_ymap}
 D^{\gamma j, j}_{jm jn} (g) \hspace{1em} & = \hspace{1em} \raisebox{-0.45\height}{\includegraphics[scale=1]{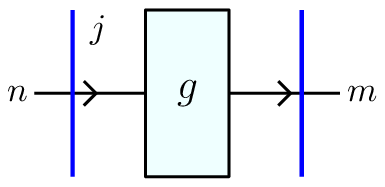}} \ .
\end{align}
\new{If we apply the $Y_\gamma$ map \eqref{eq:graphical_ymap} to the product $g_1 g_2$ and use the decomposition \eqref{eq:irrepsl2ccomp}, in the graphical notation we have one blue line at both ends:}
\begin{align}
\label{eq:graphical_ymap_g1_g2}
 D^{\gamma j, j}_{jm jn} (g_1 g_2) \hspace{1em} & = \hspace{1em} \raisebox{-0.45\height}{\includegraphics[scale=1]{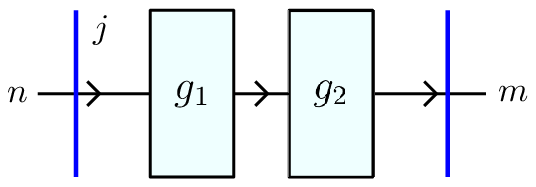}} \ .    
\end{align}
\new{The (infinite) sum over two pairs of magnetic indices is implied in graphical notation, according to equation \eqref{eq:irrepsl2ccomp}}.
If we contract two representation lines with a $Y_\gamma$ map we only sum over \new{one pair of} magnetic indices:
\begin{align}
\label{eq:ymapcontraction}
 \sum_{|p|\leq j} D^{\gamma j, j}_{jm jp} (g_2) D^{\gamma j, j}_{jp jn} (g_1) \hspace{1em} & = \hspace{1em} \raisebox{-0.45\height}{\includegraphics[scale=1]{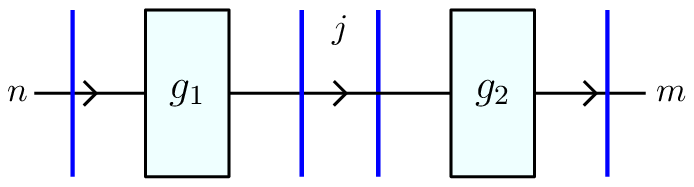}} \ .    
\end{align}
We denote with a thicker red line the sum over the spin associated to that representation $j$ weighted by a dimensional factor $(2 j +1)$: 
\begin{align}
\label{eq:implicit_sum_with_loop}
    \sum_{j} (2j+1) \sum_{m,n} D^{\gamma j, j}_{j m j n}(g_{1})D^{\gamma j, j}_{j n j m}(g_{2})\hspace{1em} & = \hspace{1em} \raisebox{-0.45\height}{\includegraphics[scale=1]{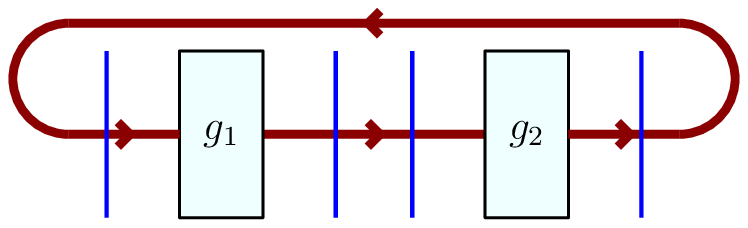}} \ .
\end{align}
The (tensor) product of two representations is represented as two lines side by side. If the group element is the same we use a single box. Similarly for the Y-map, we use a single line. When we draw a box in amplitudes, we will always imply the integration with the Haar measure over the corresponding \sltc{} group element: 
\begin{align}
    \int \dd g 
    D^{\gamma j_1, j_1}_{j_1 m_1 j_1 n_1}(g)
    D^{\gamma j_2, j_2}_{j_2 m_2 j_2 n_2}(g)
    D^{\gamma j_3, j_3}_{j_3 m_3 j_3 n_3}(g)
    D^{\gamma j_4, j_4}_{j_4 m_4 j_4 n_4}(g)
    \hspace{1em} & = \hspace{1em}\raisebox{-0.5\height}{\includegraphics[scale=1]{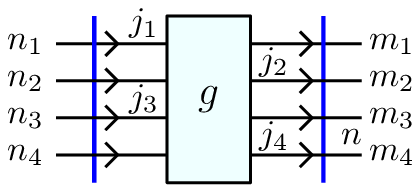}} \ .    
\end{align}

\new{\subsection{An example: writing the $\Delta_4$ amplitude}}
With the general recipe discussed in this Section and the corresponding graphical representation, we write the $\Delta_4$ spin foam amplitude associated with the 2-complex in Figure \ref{fig:2skeleton}. We also inherit the naming convention from the 2-complex. We report the amplitude in Equation~\ref{eq:Delta4Amplitude}.

To assign a unique name to all the \sltc{} group elements, we denoted as $g$ and $\tilde{g}$ the two group elements associated with the same (bulk) edge but belonging to different vertices. 
We used a small abuse of notation in writing \eqref{eq:Delta4Amplitude}. Some group elements appears as their inverse. To represent them as a single box we opted to not distinguish them. However, following our conventions, the group element in the matrix element of a target half-edge appears always as its inverse. For example, the half edge $1234$ is the source of $234$ and the target of $134$. The group element $g_{1234}$ appears as $D^{\gamma j_{234}, j_{234}}(g_{1234})$ and $D^{\gamma j_{134}, j_{134}}(g_{1234}^{-1})$. 

As mentioned \new{above}, we contracted all the boundary magnetic indices with four valent intertwiners (12 in total) as part of the prescription of the boundary data. We chose the same recoupling basis on each of them and kept the label generic for the moment ($i_e$ with $e$ a quadruple identifying a boundary tetrahedron). 

We highlighted in red the bulk face $(123)$, dual to the triangle $123$ in the $\Delta_4$ triangulation \ref{fig:geometry}. \new{According to equation \eqref{eq:implicit_sum_with_loop}}, we are implying a summation over the spin $j_{123}$ assigned to it weighted by a dimensional factor $2j_{123}+1$. As part of the boundary data, we also prescribed all the spins associated with the boundary faces. We keep them generic for the moment ($j_f$ with $f$ a triple identifying a boundary triangle). 

We regularized the amplitude removing one \sltc{} integration for each vertex as discussed \new{above}. In \eqref{eq:Delta4Amplitude} we indicate the removed integrals with a white box. This choice is arbitrary, and the amplitude value is independent of this choice. However, we can use this arbitrariness to simplify the numerical computation (see Section~\ref{sec:numerical}) by making the symmetric choice. The integral removed is always opposite to the two bulk half edges and the bulk edge $(123)$.

\begin{equation}
    \label{eq:Delta4Amplitude}
     A_{\Delta_4} = \hspace{1em} \raisebox{-0.5\height}{\includegraphics[width=0.85\textwidth]{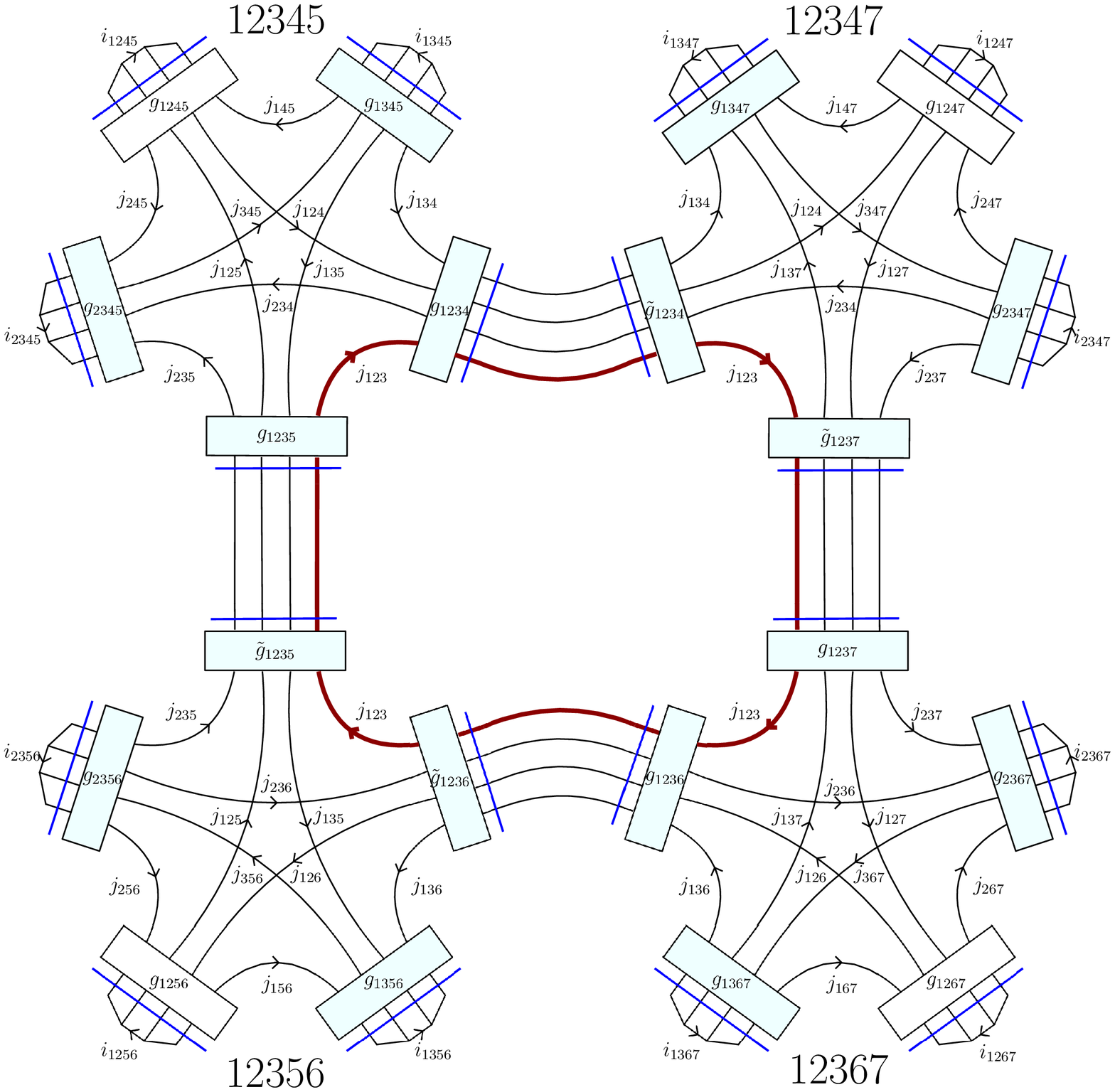}}
\end{equation}

\section{How-To divide the EPRL transition amplitudes}
\label{sec:manyvertices}
%
\new{Approaching the calculation of the full amplitude is an arduous task. The group matrix elements in unitary representations are highly oscillating functions. The integrals are group integrals over many copies (sixteen in our example) of six-dimensional non-compact groups. We divide the transition amplitude into smaller and more manageable components and compute them serialized. This approach is the most advantageous if your goal is to obtain a number from the computation of a transition amplitude. However, this could be suboptimal for semiclassical calculation due to the number of components.
}

Without any loss of generality, we insert a resolution of the identity over the intertwiner space between every two vertices of \eqref{eq:Delta4Amplitude}. 
\begin{equation}
  \label{eq:resolution1}
  \raisebox{-0.5\height}{\includegraphics[scale=0.8]{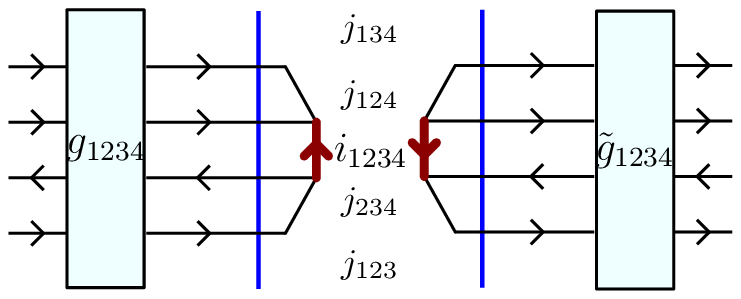}} \ .   
\end{equation}
We rewrite the resolution of the identity over the intertwiner space as an integral over \SUT{} of four matrix elements \eqref{eq:fourwigner}. We commute the \SUT{} integral with the $Y_\gamma$ map and bring the \SUT{} group element in the \sltc{} representation. 
\begin{equation}
  \label{eq:resolution2}
  \raisebox{-0.5\height}{\includegraphics[scale=0.8]{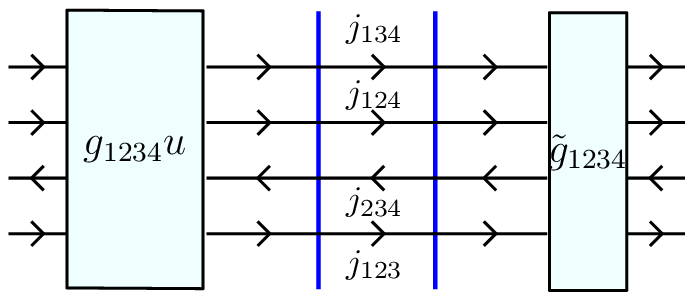}} \ . 
\end{equation}
Finally, we use the invariance property of the \sltc{} Haar measure to reabsorb the \SUT{} group element with a change of variable, obtaining the original spin foam edge. 
\begin{equation}
  \label{eq:resolution3}
  \raisebox{-0.5\height}{\includegraphics[scale=0.8]{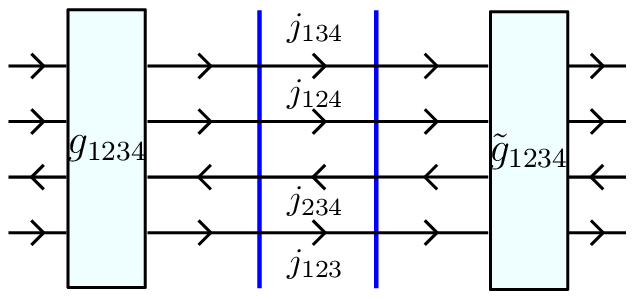}} \ .   
\end{equation}

%
\new{\subsection{An example: decomposing the $\Delta_4$ amplitude}}
\new{If we divide the $\Delta_4$ amplitude \eqref{eq:Delta4Amplitude} inserting 4 resolutions of the identity (each one between two different vertices), the latter} decomposes into a linear combination of the product of four amplitudes. That is, one per vertex. These amplitudes are commonly known as \textit{vertex amplitudes}.
\new{Using the graphical representation, we write the full $\Delta_4$ transition amplitude as:}
\begin{equation}
\label{eq:amplitude_d4final}
  A_{\Delta_4} = \sum_{l_f}\raisebox{-0.5\height}{\includegraphics[width=0.8\textwidth]{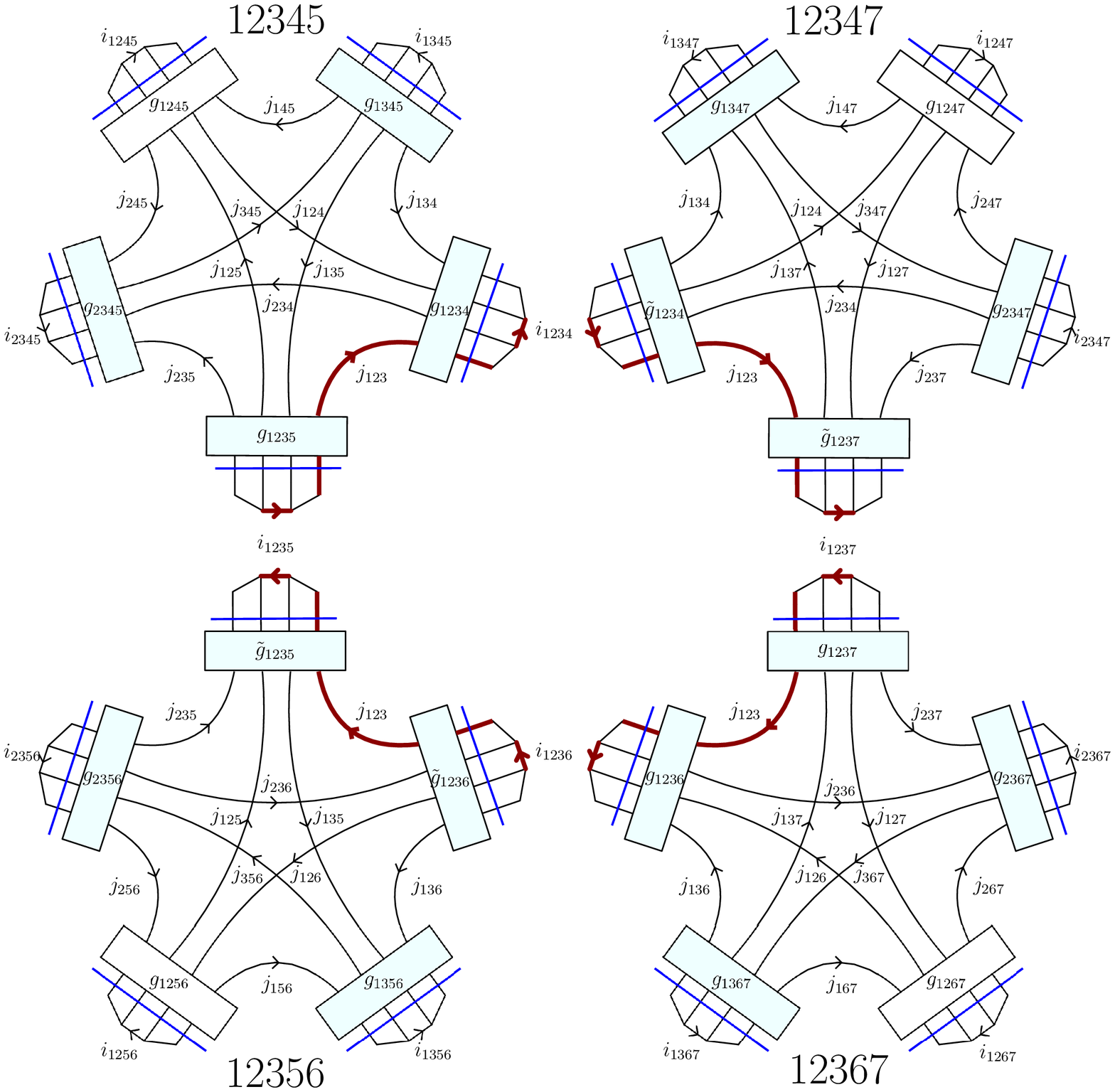}} \ .
\end{equation}
\new{By separating the vertices as in \eqref{eq:amplitude_d4final}, we have transformed the problem of calculating the full amplitude into the computation of the single building blocks: the vertex amplitudes.}

\section{How-To compute the EPRL vertex amplitudes}
\label{sec:single_vertex}

In this section, we will focus on contributions local at the vertices. For concreteness, we model the definition of the vertex amplitude on the $(12345)$ vertex in the example \eqref{eq:Delta4Amplitude}. 
\begin{equation}
  \label{eq:vertexamplitude}
A_{v_{12345}}=\hspace{1em} \raisebox{-0.5\height}{\includegraphics[scale=0.8]{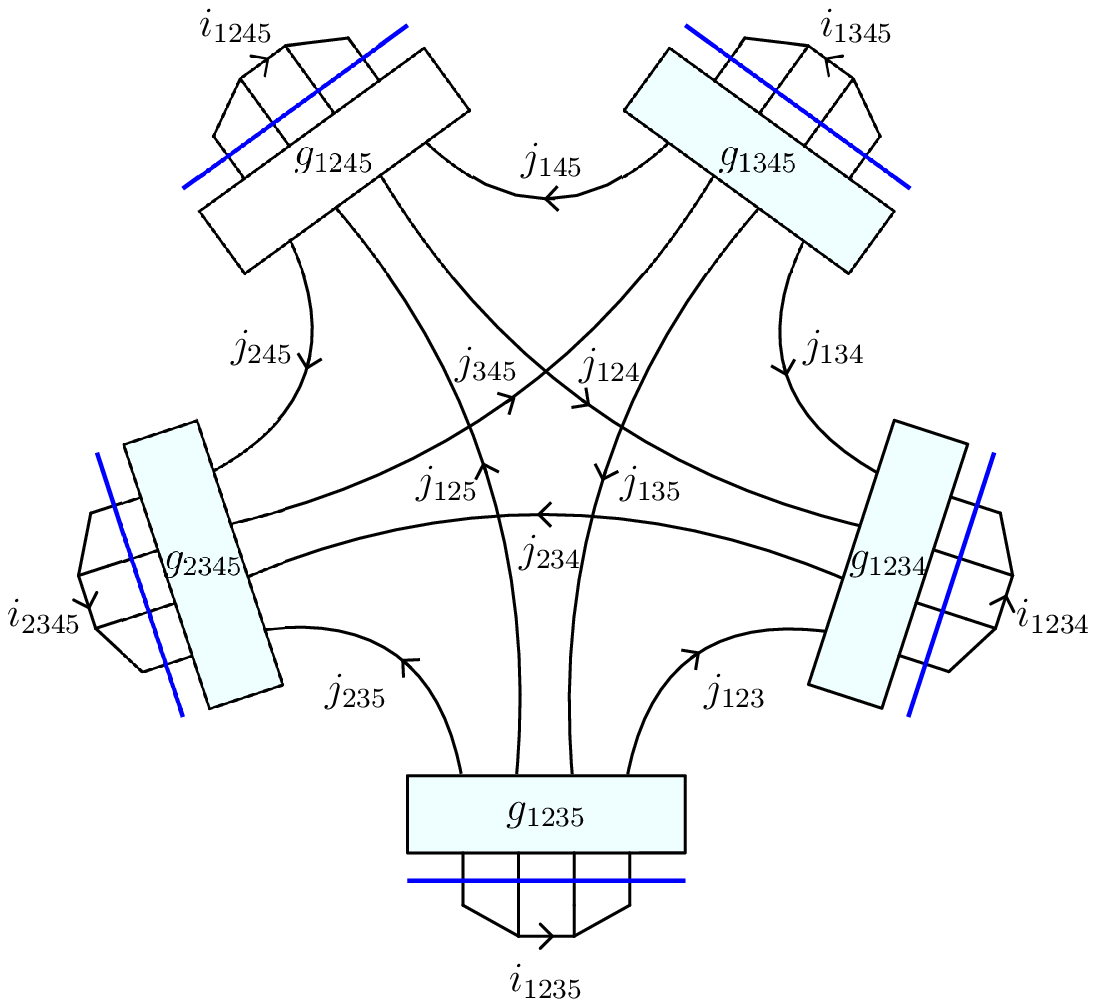}} \ .   
\end{equation}

In \eqref{eq:vertexamplitude} we contracted the magnetic indices of the bulk edges $(1234)$ and $(1235)$ with two intertwiners, labelled by $i_{1234}$ and  $i_{1235}$. We will see in the next section why this choice is natural, and we are not losing any generality. Remember that we regularized the amplitude by fixing the group element $g_{1245}=\mathds{1}$, graphically denoting such element by leaving it blank.

Consider the contribution from the wedge $(234)$. We use the representation property to separate the matrix elements corresponding to the two group elements.
\begin{align}
  \label{eq:Wigner_sl2c_basic_def}
  D^{\gamma j_{234},j_{234}}_{j_{234} m'_{234} j_{234} n_{234}}(g_{2345}^{-1} g_{1234}) =& 
  \sum_{|l_{234}|\geq j_{234}}\sum_{|n_{234}|\leq l_{234}} D^{\gamma j_{234},j_{234}}_{j_{234} m'_{234} l_{234} n_{234}}(g_{2345}^{-1}) D^{\gamma j_{234},j_{234}}_{l_{234} n_{234} j_{234} m_{234} }(g_{1234}) \\
  =& \hspace{1em} \raisebox{-0.5\height}{\includegraphics[scale=1]{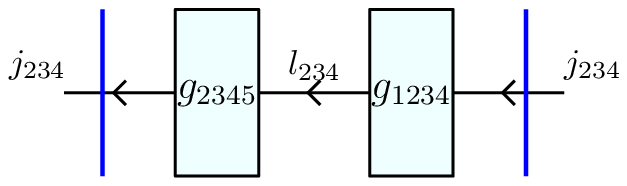}} \ .    \nonumber
\end{align}
The inverse $g_{2345}$ is due to the orientation of the wedge $(234)$ and the conventions we are adopting.

To help the reader remember about the extra summation introduced by the representation property, we wrote spin $l_{234}$ even if we are summing over it. This summation is bounded from below by $j_{234}$ but is unbounded from above. It is a consequence of the non-compactness of the group (all unitary irreducible representations are infinite-dimensional). Each group element appears as the argument of four matrix elements. For example, $g_{1234}$ appears in the matrix elements
\begin{equation}
  \label{eq:integral}
\raisebox{-0.5\height}{\includegraphics[scale=1]{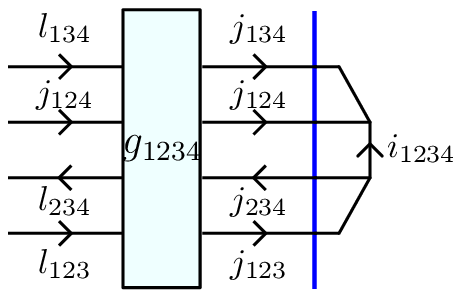}} \ .    
\end{equation}
On the face $(124)$ there is no sum over the spin $l_{124}$, as a consequence of the regularization choice $g_{1245}=\mathds{1}$ and the presence of the $Y_\gamma$ map on the half-edge $(1245)$. 

We parametrize each Lorentz transformation (\sltc{} group element) with an arbitrary rotation (\SUT{} group element) followed by a boost in a conventional direction (the $3$ direction in our case) and another arbitrary rotation: the Cartan parametrization \eqref{eq:CartanSL2C} of \sltc. The representation matrices decompose as \eqref{eq:CartanSL2C} and the Haar measure factorizes as in \eqref{eq:HaarSL2C}. We divide the contribution of the integral on the half-edge $(1234)$ to the amplitude into 
\begin{equation}
  \label{eq:cartanhalfedge}
\raisebox{-0.5\height}{\includegraphics[scale=1]{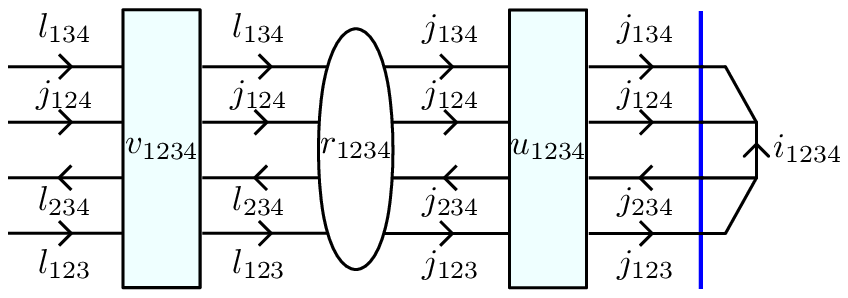}} \ .    
\end{equation}

The matrix elements of $u_{1234}$ and $v_{1234}$ are \SUT{} matrix elements elements \eqref{eq:sltcsu2}. We represent the integral over the rapidity $r_{1234}$ of the product of four reduced matrix elements \eqref{eq:dSL2C} as a white oval. We wrote the arguments explicitly to help the reader to visualize the parametrization. In the following, we will omit the name of redundant integration variables. 

The $Y$ map commutes with \SUT{} group elements. Therefore, we move it next to the rapidity integral. We perform the integrals over \SUT{} \eqref{eq:fourwignerintegral} in terms of $(4jm)$ symbols. The contribution to the amplitude from the half-edge \eqref{eq:cartanhalfedge} is
\begin{equation}
  \label{eq:booster1}
\raisebox{-0.5\height}{\includegraphics[scale=1]{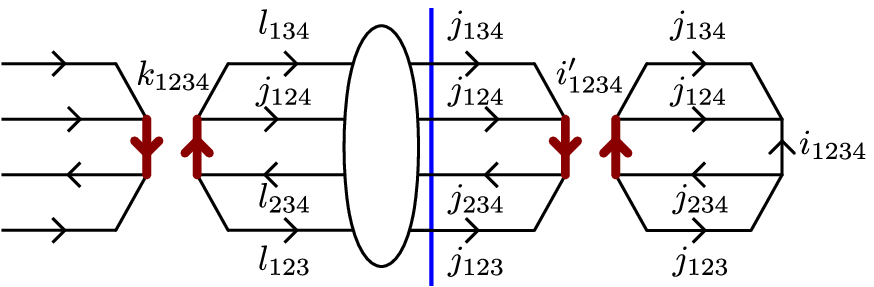}} \ ,
\end{equation}
where the thicker red line represent a summation over the corresponding label weighted by a dimensional factor as in \eqref{eq:implicit_sum_with_loop}. 

The contraction of two $(4jm)$ symbols obey the orthogonality condition \eqref{eq:4jmtheta} and allows us to remove the summation over $i'_{1234}$ 
\begin{equation}
  \label{eq:thetha4_pdona}
\raisebox{-0.45\height}{\includegraphics[scale=1]{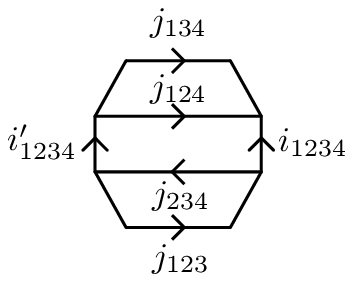}} \hspace{1em} = \hspace{1em} (-1)^{2j_{234}}\frac{\delta_{i'_{1234}i_{1234}}}{2 i_{1234}+1}\ .
\end{equation}
where the phase $(-1)^{2j_{234}}$ is a consequence of the different orientation of the link \eqref{eq:lineinversion}.

We define the booster functions $B_4^\gamma$ as the result of the integral
\begin{align}
  \label{eq:boosterdef}
  B_4^\gamma &\left( j_1,j_2,j_3,j_4, l_1,l_2,l_3,l_4 ; i,k\right) 
  =  \raisebox{-0.5\height}{\includegraphics[scale=1]{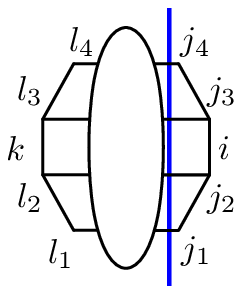}} =  \\
  &=\sum_{ p_1, p_2, p_3, p_4 } 
  \left(\begin{array}{cccc} l_1 & l_2 & l_3 & l_4 \\ p_1 & p_2 & p_3 & p_4 \end{array}\right)^{(k)}
  \left(\int_0^\infty \dd r \frac{1}{4\pi}\sinh^2r \, \bigotimes_{f=1}^4 d^{\gamma j_f,j_f}_{l_f j_f p_f}(r) \right)
  \left(\begin{array}{cccc} j_1 & j_2 & j_3 & j_4 \\ p_1 & p_2 & p_3 & p_4 \end{array}\right)^{(i)} 
  \ . \nonumber
\end{align}
The booster functions were first introduced in \cite{Speziale2016}, numerically computed in \cite{Dona2018,Francesco_draft_new_code}, analytically evaluated in terms of complex gamma functions  \cite{Anderson:1970ez, Kerimov:1978wf}, and they have an interesting geometrical interpretation in terms of boosted tetrahedra \cite{Dona:2020xzv}. The booster functions encode how the EPRL model imposes the quantum simplicity constraints and depend on the Immirzi parameter $\gamma$. Note that, in the definition \eqref{eq:boosterdef}, we dropped the information on the orientation of the faces. The orientation of the faces in the booster function is irrelevant. The effect of orientation change of the $(4jm)$ symbols cancels exactly the effect of orientation change of the reduced density matrices of \sltc, as we discuss in Appendix \ref{app:sltc}. Using this definition, we can write \eqref{eq:integral} in terms of the booster functions as:

\begin{align}
\label{eq:halfedgebooster}
\raisebox{-0.5\height}{\includegraphics[scale=1]{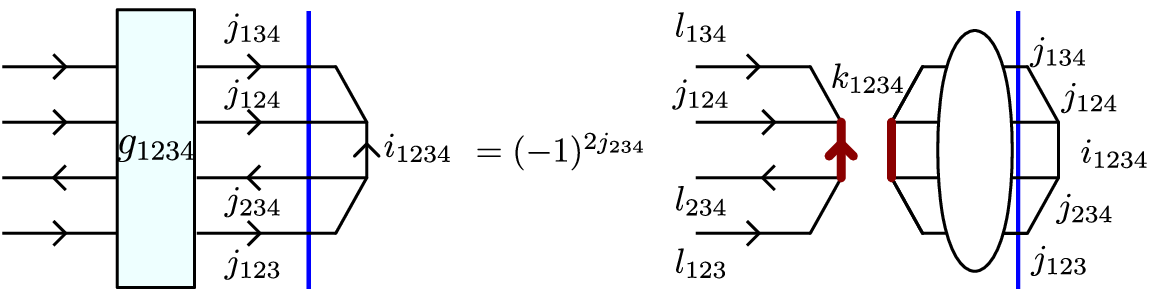}} \ .
\end{align}

We compute the contribution to the amplitude from all other half edges of $(12345)$ with the same prescription. The $(4jm)$ symbols in \eqref{eq:halfedgebooster} contracts among themselves and form a $\{15j\}$ symbol of the first kind \eqref{eq:15jsymbol}. 
\begin{align}
  \label{eq:vertex_amplitude_with_boosters}
A_{v_{12345}}= (-1)^{2 j_{135} + 2 j_{234}}\sum_{l_f} \raisebox{-0.5\height}{\includegraphics[scale=0.75]{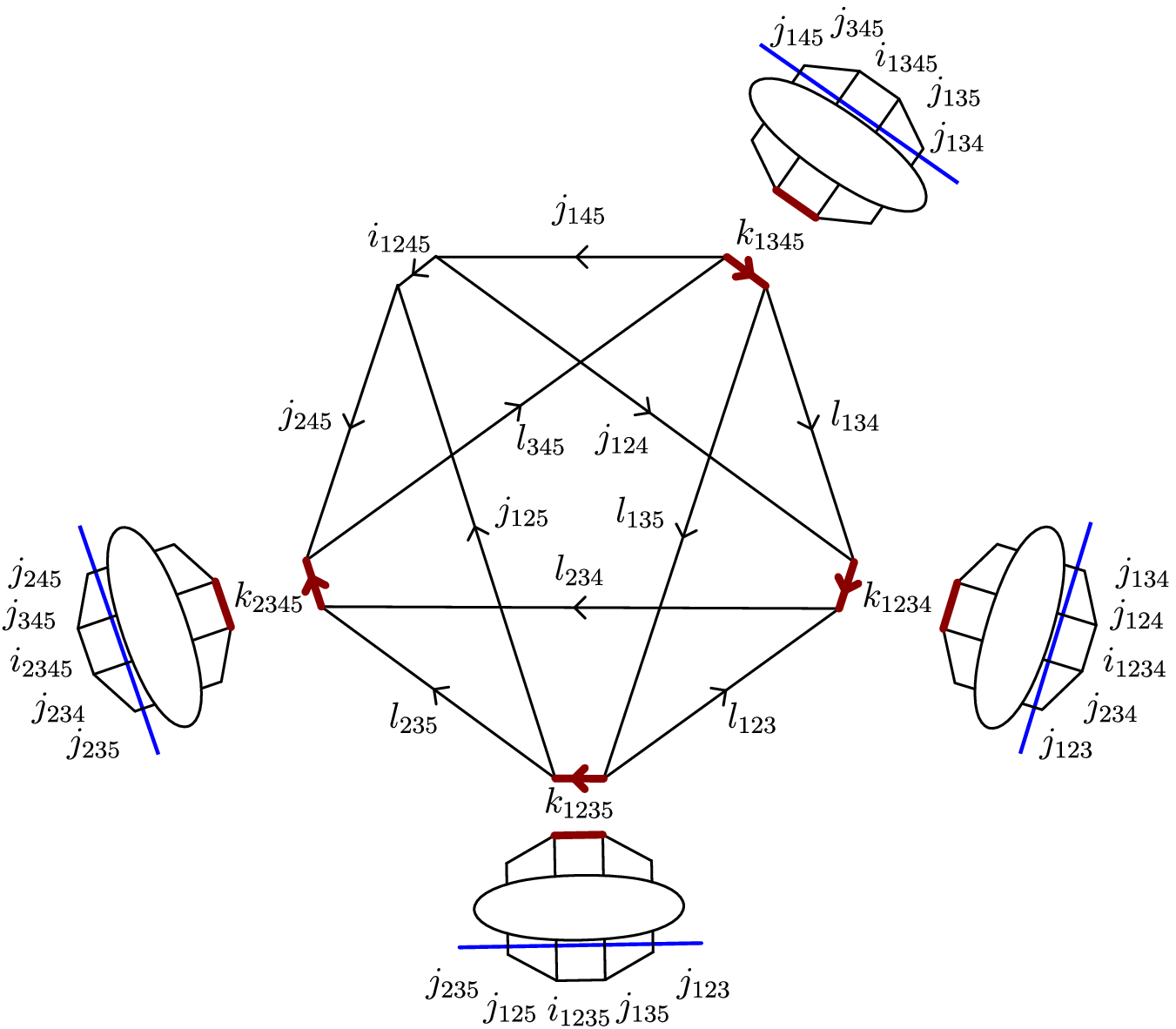}} \ .
\end{align}
The sum over spins $l_f$ are only bounded from below (e.g. $l_{123}\geq j_{123}$) and the intertwiners $k_e$ are bounded by the triangular inequalities of the $(4jm)$ symbols. To complete the calculation we recognize the \SUT invariant as a canonical $\{15j\}$ symbol of the first kind \eqref{eq:15jsymbol}. The vertex amplitude is
\begin{align}
\label{eq:amplitude-formula}
A_{v_{12345}}= (-1)^{2 j_{135} + 2 j_{234}} \sum_{l_f} &\left\lbrace \begin{array}{ccccc} i_{1245} & j_{124} & k_{1234} & l_{234} & k_{2345} \\
j_{145} & l_{134} & l_{123} & l_{235} & j_{245} \\ 
l_{345} & k_{1345} & l_{135} & k_{1235} & j_{125} \end{array}\right\rbrace \\
 &B_4^\gamma\left(j_{235} , j_{234}, j_{345}, j_{245}, l_{235}, l_{234}, l_{345}, j_{245} ; i_{2345},k_{2345}\right) \nonumber \\
 &B_4^\gamma\left(j_{123} , j_{135}, j_{125}, j_{235}, l_{123}, l_{135}, j_{125}, l_{235} ; i_{1235},k_{1235}\right) \nonumber \\
 &B_4^\gamma\left(j_{134} , j_{124}, j_{234}, j_{123}, l_{134}, j_{124}, l_{234}, l_{123} ; i_{1234},k_{1234}\right) \nonumber \\
 &B_4^\gamma\left(j_{145} , j_{345}, j_{135}, j_{134}, j_{145}, l_{345}, l_{135}, l_{134} ; i_{1345},k_{1345}\right) \ .\nonumber
\end{align}
In general one need to change the orientation of some links to obtain the canonical $\{15j\}$ symbol using \eqref{eq:lineinversion} to compute the relative phase. 

\medskip

We rewrote the vertex amplitude \eqref{eq:vertexamplitude} as a combination of a canonical $\{15j\}$ symbol weighted by four booster functions. 


\section{How-To calculate numbers}
\label{sec:numerical}
In the previous Section, we completed the formal evaluation of the amplitude. If we are satisfied with the expression \eqref{eq:amplitude_d4final} we can stop here. A few more steps are needed if we want to translate it into a number.
\new{We decompose each vertex amplitude in \eqref{eq:amplitude_d4final} as in \eqref{eq:vertex_amplitude_with_boosters}. By doing so, we finally write the $\Delta_4$ transition amplitude in the appropriate form for a numerical evaluation:}
\begin{equation}
\label{eq:amplitude_d4final_boosters}
  A_{\Delta_4} = \sum_{l_f}\raisebox{-0.5\height}{\includegraphics[width=0.8\textwidth]{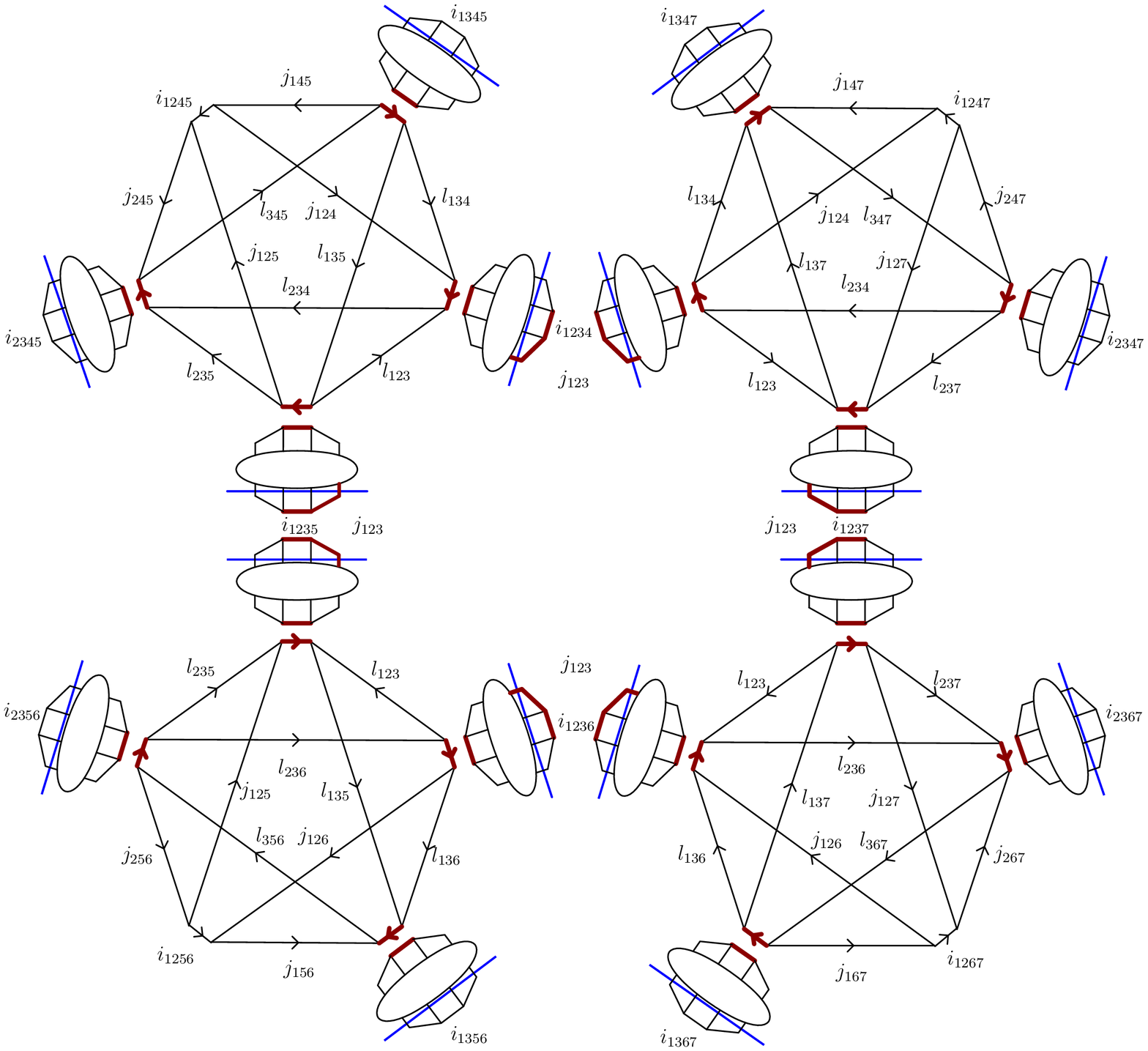}} \ .
\end{equation}
In this paper, we rely on the library \texttt{sl2cfoam-next} to perform the numerical evaluation of the EPRL spin foam amplitude. The code discussed in this Section is available in the repository in the form of notebooks \cite{Paper_repository}.
\new{\subsection{Historical overview}}
The development of a library for the numerical computation of the Lorentzian EPRL 4-simplex vertex amplitude started with \texttt{sl2cfoam} \cite{old_sl2cfoam}. The library is coded in \texttt{C} and is based on the decomposition of the vertex amplitude \eqref{eq:amplitude-formula} in terms of booster functions. We refer to the original paper \cite{Dona2018} for a detailed discussion of the library's performances, accuracy, and memory management. 

The library computes the \SUT{} invariant symbols using \texttt{wigxjpf} \cite{Wigxjpf_library}. The invariants are stored efficiently in custom hash tables based on \texttt{khash} \cite{Klib_library} that takes into account their symmetry properties. 

\texttt{sl2cfoam} computes the booster functions performing a numerical integration of the boost matrix elements \eqref{eq:boosterdef}. The integrand is rewritten as a finite sum of exponentials with complex coefficients to tame its highly oscillating behaviour. One obtains the booster function from the interference of many exponential integrals done with the trapezoidal rule. In order to reach enough numerical precision, the authors employed arbitrary precision floating point computations with the GNU libraries \texttt{GMP} \cite{GMP_library}, \texttt{MPFR} \cite{MPFR_library} and \texttt{MPC} \cite{MPC_library}. 

The library was used to explore the numerical properties of the EPRL vertex \cite{Dona:2020tvv,Dona2019, Dona:2018infrared, Sarno2018}. The need for a much more efficient and accurate code immediately became clear, as the computational time for more complex amplitudes was definitely out of reach. 

Recently \texttt{sl2cfoam-next}\cite{sl2cfoam_next}, the evolution of \texttt{sl2cfoam}, has been released. Also the new library is written in \texttt{C}, but it has an optional \texttt{julia} interface \cite{bezanson2017julia} which hugely simplifies its usage. Although \texttt{sl2cfoam-next} computes the Lorentzian EPRL vertex amplitude in the form \eqref{eq:amplitude-formula}, it introduces several ideas and techniques borrowed from High-Performance Computing and tensor networks. Therefore, with respect to the original version, it represents a significant improvement in performance and precision. We refer to \cite{Francesco_draft_new_code} for its complete description and several usage examples.


The numerical integration of the booster functions is performed with the Gauss-Kronrod quadrature method after a weighted sub-intervals decomposition of the integration range. Also, for technical reasons, the $\gamma$-simple unitary irreducible representations in the principal series of \sltc{} are slightly different from \eqref{eq:ymap} as it uses $D^{\gamma(j+1) j}$ instead of $D^{\gamma j j}$. 

The huge number of sums and products involved in the expression \eqref{eq:vertexamplitude} is performed with optimized routines for multidimensional arrays multiplications (we will refer to them loosely as tensors in the informatics sense), such as BLAS \cite{blackford2002updated} and MKL. For the description of the CPU parallelization scheme adopted, we refer to \cite{Francesco_draft_new_code}. It has been recently introduced the possibility to offload tensor contractions to the GPU and parallelize them over the GPU cores with the \texttt{CUDA} platform \cite{cuda} by using the \texttt{julia} package \texttt{CUDA.jl} \cite{besard2018juliagpu, besard2019prototyping}.

\new{\subsection{Introducing the cut-off}}

The vertex amplitude \eqref{eq:amplitude-formula} is made of three distinct elements: the $\{15j\}$ symbol, the booster functions and the combination of two together. The formula \eqref{eq:amplitude-formula} is exact and \texttt{sl2cfoam-next} can compute its constituents to very high numerical precision. However, the sums over the spins $l_f$, that appear in \eqref{eq:amplitude-formula} due to the split of the representation matrix elements on the wedges, are bounded from below but not from above. This means that in order to extract a number from \eqref{eq:amplitude-formula} we need to make an approximation and cut-off the 6 unbounded sums in the vertex amplitude. 
While unbounded the sums are convergent because the vertex amplitude is finite \cite{Engle:2008ev}. Therefore, in principle, it is possible to find a cut-off large enough to capture the value of the amplitude with the desired precision. 

The library \texttt{sl2cfoam-next} implement an homogeneous cut-off $\Delta l$ on all the unbounded summations. We replace the sums 
\begin{equation}
    \sum_{l_f = j_f}^\infty \quad \longrightarrow \quad \sum_{l_f = j_f}^{j_f + \Delta l} \ .
\end{equation}

Unfortunately, we do not have a prescription to find the optimal value of $\Delta l$. Numerical explorations show that it depends on the details of boundary data, such as the face spins $j_f$ and the Barbero-Immirzi parameter $\gamma$. At the moment, the best consolidated strategy is to set $\Delta l$ as large as possible and estimate the error by studying the value of the amplitude $A_{\Delta_4}(\Delta l)$ as a function of the cut-off. 

Recently \cite{frisoni2021numerical} introduced an extrapolation scheme to overcome the enormous computational cost represented by indefinitely increasing $\Delta l$. The extrapolation algorithm was used to calculate the self-energy spinfoam amplitude (see \cite{Riello:2013bzw, Dona:2018infrared}), which is a divergent amplitude since it contains a bubble. Furthermore, we mention that the implementation of Markov Chain Monte Carlo methods in the study of spinfoams based on the techniques discussed in this paper is in progress \cite{Cosm_project}. \\
\new{\subsection{Using the \texttt{sl2cfoam-next}}} 
Before any calculation we need to import the \texttt{sl2cfoam-next} library and initialize it. In the blocks of code of this Section, we will imply that the library is correctly initialized first, and we omit the following code.
\begin{jllisting}[caption={Initialization of \texttt{sl2cfoam-next}},label={code-init}]
using SL2Cfoam
Immirzi = 1.2
sl2c_data_folder = "$(path_to_library_data_folder)"
sl2c_configuration = SL2Cfoam.Config(VerbosityOff, VeryHighAccuracy, 100, 0)
SL2Cfoam.cinit(sl2c_data_folder, Immirzi, sl2c_configuration) 
\end{jllisting}
We set the value of the Immirzi parameter to $1.2$ (for historical reasons, any value is equally possible). We define a data folder that is used both to look for the \texttt{fastwigxj} tables and to store the computed data optionally. In this way, we avoid recomputing the same vertex amplitude for a second time. We refer to the documentation of \texttt{sl2cfoam-next} and the accompanying paper \cite{Francesco_draft_new_code} for a detailed description of all the setup options. 

\subsection{Computing one vertex} 

We find very valuable to dedicate this paragraph to show how to use the \texttt{julia} frontend of \texttt{sl2cfoam-next} to compute the EPRL vertex amplitude. We use the amplitude \eqref{eq:amplitude-formula} as reference. We provide some \texttt{jupyter} notebooks\footnote{The code in \cite{Paper_repository} was tested with the kernel \texttt{julia 1.7.0}} in the Git repository \cite{Paper_repository}, for interactive usage examples that the reader can compile and execute. In the Listing~\ref{code1} we show an essential schematic representation of the code.
\begin{jllisting}[caption={Computation of a vertex amplitude with \texttt{sl2cfoam-next}},label={code1}]
Dl = 15
spins = j245, j125, j124, j145, j235, j234, j345, j123, j135, j134 = ones(10)
@time v = vertex_compute(spins, Dl);
\end{jllisting}
We are omitting the initialization code in Listing~\ref{code-init}.
In lines 1-2, we specify the boundary data (all the spins equal to 1) and the cut-off $\Delta l=15$. In line 3, we compute the amplitude. The function \texttt{vertex\_compute} returns a tensor with five indices, one per intertwiner, computing the vertex amplitude \eqref{eq:vertex_amplitude_with_boosters} (without any phase) for all possible values of boundary intertwiners. In \cite{Paper_repository} we show how to compute a restricted range of boundary intertwiners. 

The \texttt{@time} macro is used for logging purposes, tracking the computational time and memory usage. At fixed boundary spins and Immirzi parameter, the computation time depends on several parameters such as the value of the cut-off $\Delta l$ and the accuracy level at which the library is set. With the parameters specified in Listing~\ref{code-init}, the first time that line 3 of Listing~\ref{code1} is run takes 158 seconds. The computation time decreases exponentially by selecting a lower cut-off $\Delta l$. We tested this code on a laptop with  Intel(R) Core(TM) i7-10750H 2.60GHz processor. The library distributes the workload on the available cores, according to the parallelization scheme discussed in \cite{Francesco_draft_new_code}. If we store the required booster functions during the first computation, the second time we run the script takes 3.2 seconds. It is the time to compute, sum, and contract all the required $\{ 15j \}$ symbols in the expression \eqref{eq:vertex_amplitude_with_boosters}. Finally, If we store the full vertex amplitude, the computation time is negligible since nothing is calculated, and we retrieve the value from memory.  

If we are interested in one single vertex amplitude this is all we need to do. 

\new{\subsection{An example: computing the $\Delta_4$ amplitude with \texttt{sl2cfoam-next}}} 

We split the computation of the amplitude \eqref{eq:amplitude_d4final_boosters} into two steps. First, we compute and save the value of all the necessary vertices. Then, we contract the required vertices to calculate the $\Delta_4$ amplitude. For simplicity, we fix all boundary spins $j$ equal to $1$. The bulk spin $j_{123}$ assumes values from $0$ to $3j$, while bulk intertwiners $i_{1234}$, $i_{1235}$, $i_{1236}$, and $i_{1237}$ assume values compatible with triangular inequalities. With the regularization choices we did, the vertex amplitudes are fully symmetric. That is, the bulk spin and bulk intertwiners always appear in the same position in each of the four vertices. Therefore, it is sufficient to compute only a single vertex amplitude for all the possible values of spins and intertwiners. To keep the computational time reasonable, we fix the cut-off $\Delta l$ to $15$. We analyze the dependence of the amplitude on this cut-off in the next step.
\begin{jllisting}[caption={Computation of all the vertex amplitudes needed in the computation of the transition amplitude \eqref{eq:amplitude_d4final_boosters}},label={code2}]
using JLD2 
j = 1
Dl = 15
root_dir = pwd()
vertex_path = "$(root_dir)/vertex_ampls/Immirzi_$(Immirzi)/j_$(j)/Dl_$(Dl)" 
mkpath(vertex_path)
j_bulk_min, j_bulk_max = 0, 3j
for j_bulk = j_bulk_min:j_bulk_max
    spins = [j, j, j, j, j, j, j, j_bulk, j, j]
    v = vertex_compute(spins, Dl)        
    vertex = v.a        
    @save "$(vertex_path)/j_bulk_$(j_bulk)_fulltensor.jld2" vertex    
end
\end{jllisting}
In lines 2-3, we set all boundary spins equal to 1 and the cut-off $\Delta l =15$. In lines 4-6, we create the directory path to organize the files containing the computed amplitudes. In line 7, we define the range of the bulk spin, and from line 8, we loop over it. In lines 9-12, we assign the vertex amplitude's spins, compute the vertex amplitude, and save it for later use. Notice that we are computing the \textit{fulltensor} vertex amplitude, namely for all the possible values of boundary intertwiners. This ensures that the $\Delta_4$ amplitude can be calculated for any combination of the latter.

Finally we compute the whole amplitude \eqref{eq:amplitude_d4final_boosters} by assembling all the vertices. One of the main advantages of collecting the vertex amplitudes in multidimensional arrays is that there are very efficient methods to multiply (or ``contract'') the latter. For the application we discuss in this work it is unnecessary to improve upon a \texttt{for} loop, but \texttt{julia} offers the possibility to perform contractions in a wonderfully efficient and simple way, possibly using the GPU. See for example the method \texttt{contract}, provided in \texttt{sl2cfoam-next} to contract vertices with coherent boundary states. Alternatively, there are packages such as \texttt{LoopVectorization.jl} (see \cite{Star_model_repository} for an example) or libraries like \texttt{ITensor} \cite{itensor}. 
In the following block of code, we are assuming that all the variables defined in Listing~\ref{code2} are available.
\begin{jllisting}[caption={Computation of the transition amplitude \eqref{eq:amplitude_d4final_boosters}. All the vertex amplitudes are pre-computed.},label={code3}]
i_b = 2
i = i_b + 1    
D4_amp = 0.0 
for j_bulk = j_bulk_min:j_bulk_max
    fulltensor_to_load = "$(vertex_path)/j_bulk_$(j_bulk)_fulltensor.jld2"
    @load "$(fulltensor_to_load)" vertex 
    D4_partial_amp = 0.0
    D = size(vertex[i,:,:,i,i])[1]
    for i_1234 in 1:D, i_1235 in 1:D, i_1236 in 1:D, i_1237 in 1:D
        @inbounds D4_partial_amp += vertex[i,i_1234,i_1235,i,i]*vertex[i,i_1235,i_1236,i,i]*
                                    vertex[i,i_1236,i_1237,i,i]*vertex[i,i_1237,i_1234,i,i]     
    end     
    D4_partial_amp *= (2j_bulk + 1)
    D4_amp += D4_partial_amp  
end
@show D4_amp
\end{jllisting}
In lines 1-2, we define the boundary intertwiners. For simplicity, we pick them all equal to 2, but any other choice is also possible. Notice that in \texttt{julia} the vector's index starts from 1. Therefore, we shift its value. In line 3, we initialize the variable that will contain the amplitude. From line 4, we loop over all the possible values of the bulk spin. In lines 5-6, we load the precomputed amplitude. In line 7, we initialize the variable to store the partial amplitude. The partial amplitude is the quantity in \eqref{eq:amplitude_d4final_boosters} at fixed value of the bulk spin $j_{123}$. From lines 8 to 12, we sum over the bulk intertwiners the product of the four vertex amplitudes. In line 13, we add the dimensional factor to the full amplitude value, that we display in line 16.

\new{\subsection{Results and extrapolation}} 

\begin{figure}[H]
    \centering
    \includegraphics[width=0.9\textwidth]{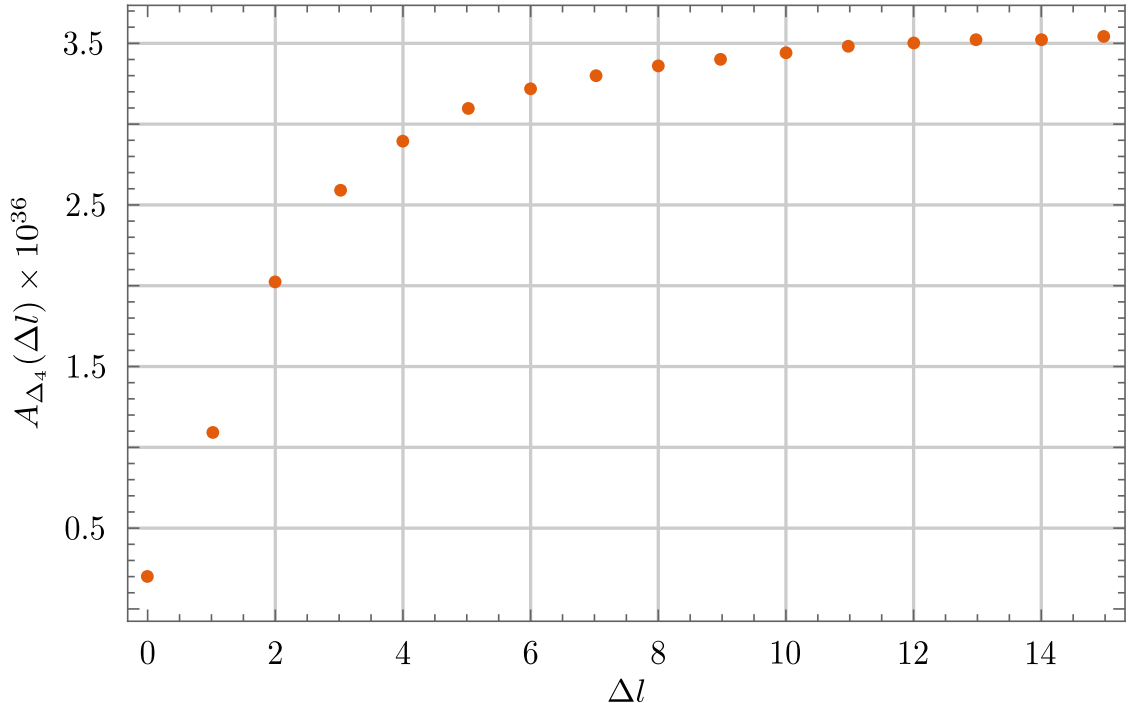}
    \caption{Amplitude $A_{\Delta_4}(\Delta l)$ in function of the cut-off.}
    \label{fig:amplitudedl15}
\end{figure}

We summarize the result of our calculation in Table~\ref{tab:amplitudedl15} and Figure~\ref{fig:amplitudedl15}.
\begin{table}[H]
    \centering
   \begin{tabular}{r|cccccccc}
  $\Delta l$    & $0$& $1$ & $2$ & $3$ & $4$ & $5$ & $6$ & $7$  \\
  $A_{\Delta_4}(\Delta l) \times 10^{36}$& $0.202$ & $1.09 $ & $2.03 $ & $2.59 $ & $2.90 $ & $3.09 $ & $3.21 $ & $3.29 $  \\
  \hline
  $\Delta l$  & $8$ & $9$ & $10$ & $11$ & $12$ & $13$ & $14$ & $15$\\
  $A_{\Delta_4}(\Delta l)$ & $3.36$ & $3.40 $ & $3.44 $ & $3.47 $ & $3.50 $ & $3.51 $ & $3.53 $ & $3.54 $
\end{tabular}
    \caption{Numerical values of the amplitude $A_{\Delta_4}(\Delta l)$ in function of the cut-off.}
    \label{tab:amplitudedl15}
\end{table}
Looking at the plot in Figure~\ref{fig:amplitudedl15} seems reasonable to deduce that by increasing the cut-off $\Delta l$, the value of the amplitude grows and (asymptotically) converges to the value of the amplitude. In the first numerical works based on \texttt{sl2cfoam} and \texttt{st2cfoam-next} \cite{Dona2019,Dona:2020tvv} the amplitude was approximated using the value with largest available cut-off. However, we have way more information (convergence, trends, speed). Is it possible to better estimate the amplitude with what we have?  \\
\new{We use series acceleration techniques. We} reorganize the sums in the amplitude such that it takes the form 
\begin{equation}
\label{eq:amplitudeasseries}
    A_{\Delta_4}(\Delta l) = \sum_{n=0}^{\Delta l} a_n \ ,
\end{equation}
where $a_0$ is the amplitude with vanishing cut-off $\Delta l = 0$ (also called \textit{simplified model} in \cite{Speziale2016}), $a_1$ encodes all the terms in the various sums of  $A_{\Delta_4}$ that appears in the amplitude cut-off $\Delta l = 1$ but are not in $a_0$, and so on. The whole amplitude is recovered in the limit for infinite cutoff of \eqref{eq:amplitudeasseries}. 

While recast in this form, we can apply techniques to estimate the value of numerical convergent series like the one in Appendix~\ref{app:seriesproof}. A similar approach was attempted in \cite{frisoni2021numerical}, and here we improve it and clarify it. \new{The technique is analog to the Aitken delta-squared process \cite{Aitken:1927} applied the the succession of the partial sum \eqref{eq:amplitudeasseries}.}

Since the amplitude is finite, the infinite cutoff limit of \eqref{eq:amplitudeasseries} exists, and the series defined in this way is convergent. We will assume that the ratios $a_n/a_{n-1}$ are increasing (from a certain point onward). This assumption is backed up by numerical evidence (up to the available cutoff). The lower bound estimate in \eqref{eq:bounds} specialized for the series \eqref{eq:amplitudeasseries} is
\begin{equation}
  \label{eq:bounds-d4-1}
 A_{\Delta_4} \gtrapprox \frac{A_{\Delta_4}(\Delta l) A_{\Delta_4}(\Delta l-2) - A_{\Delta_4}^2(\Delta l-1)}{A_{\Delta_4}(\Delta l) - 2A_{\Delta_4}(\Delta l-1)+A_{\Delta_4}(\Delta l-2)} = \frac{A_{\Delta_4}(15) A_{\Delta_4}(13) - A_{\Delta_4}^2(14)}{A_{\Delta_4}(15) - 2A_{\Delta_4}(14)+A_{\Delta_4}(13)} \approx 3.61\cdot10^{-36} \ ,
\end{equation}
where we specified the largest maximum value of the cut-off we computed, which is $\Delta l =15$. The estimate \eqref{eq:bounds-d4-1} is significantly different from $A_{\Delta_4}(15)$ and does not require any additional calculation or resources. 
\new{The lower bound \eqref{eq:bounds-d4-1} is analogous to the approximation we can obtain with the Aitken's delta-squared process.}
With \eqref{eq:bounds} we also obtain an upper bound to the amplitude.
\begin{equation}
  \label{eq:bounds-d4-2}
 A_{\Delta_4} \lessapprox \frac{A_{\Delta_4}(\Delta l) - A_{\Delta_4}(\Delta l-1) L}{1-L} = \frac{A_{\Delta_4}(15) - L A_{\Delta_4}(14)}{1-L} \approx 3.74 \cdot 10^{-36} \ ,
\end{equation}
where $L=\lim_{\Delta l \to \infty} a_n/a_{n-1}$ which we estimate numerically with a inverse power law fit as in the example in Appendix~\ref{app:seriesproof}. We stress that the validity of this upper bound needs to be taken with a grain of salt since approximating the value of $L$ can falsify the inequality in \eqref{eq:bounds-d4-2}. 
Summarizing, 
\begin{equation}
  \label{eq:bounds-d4}
  A_{\Delta_4} \in ( 3.61\cdot10^{-36} , \approx 3.74 \cdot 10^{-36}) \ .
\end{equation}
We can plot the amplitude together with the bound obtained from   \eqref{eq:bounds-d4-1} and \eqref{eq:bounds-d4-2} to appreciate the improvement to the rough estimate $A_{\Delta_4}(15)$.
\begin{figure}[H]
    \centering
    \includegraphics[width=0.9\textwidth]{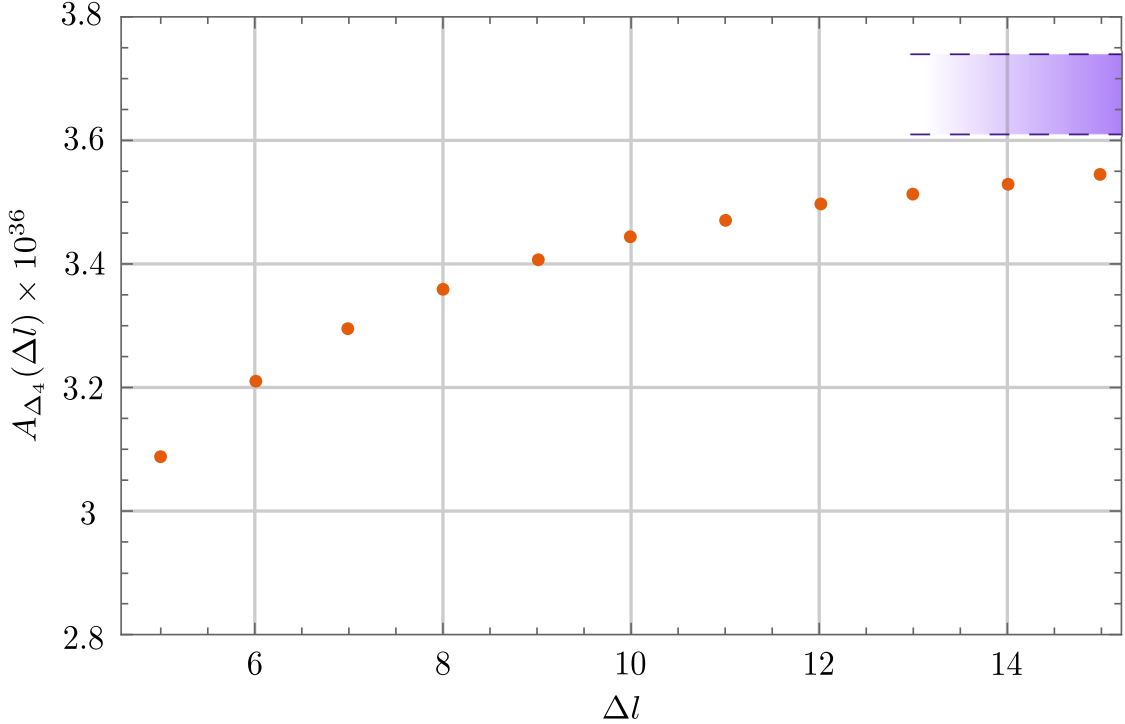}
    \caption{Amplitude $A_{\Delta_4}(\Delta l)$ in function of the cut-off with the band \eqref{eq:bounds-d4} highlighted in blue. We excluded the points $\Delta l < 5$ for a better plot scale.}
    \label{fig:amplitudedl15_first_excluded}
\end{figure}
Computational resources are precious. Up to this point, all the calculations we proposed can be done on a standard laptop. How much can we improve the estimate by increasing the cut-off using High Performance Computing? We used Compute Canada's Narval cluster to increase the cut-off from 15 to 25. The computation was distributed on 80 tasks with 10 CPUs per task, requiring about 8 minutes. The script used can be found in \cite{Paper_repository}. These were the resources we could employ in this project. There is still a big room for easy improvement.
\begin{table}[H]
    \centering
   \begin{tabular}{r|cccccccccc}
  $\Delta l$    & $16$& $17$ & $18$ & $19$ & $20$ & $21$ & $22$ & $23$ & $24$ & $25$  \\
  $A_{\Delta_4}(\Delta l)\times 10^{36}$& $3.55$ & $3.56$ & $3.57$ & $3.58$ & $3.58$ & $3.59$ & $3.59$ & $3.60$ & $3.60$ & $3.61$   \\
\end{tabular}
    \caption{Numerical values of the amplitude $A_{\Delta_4}(\Delta l)$ in function of the cut-off.}
    \label{tab:amplitudedl25}
\end{table}
Repeating the estimate process, we find new upper and lower bounds.
\begin{equation}
  \label{eq:bounds-d4-new}
 A_{\Delta_4} \in ( 3.63\cdot 10^{-36} , \approx 3.69 \cdot 10^{-36}) \ .
\end{equation}
The lower bound is marginally improved, as expected by comparing the numerical values in Table~\ref{tab:amplitudedl25} to the ones in Table~\ref{tab:amplitudedl15}. However, the improvement on the upper bound is significant. Having more points to extrapolate the limit of the ratios $L$ is essential.
\begin{figure}[H]
    \centering
    \includegraphics[width=0.9\textwidth]{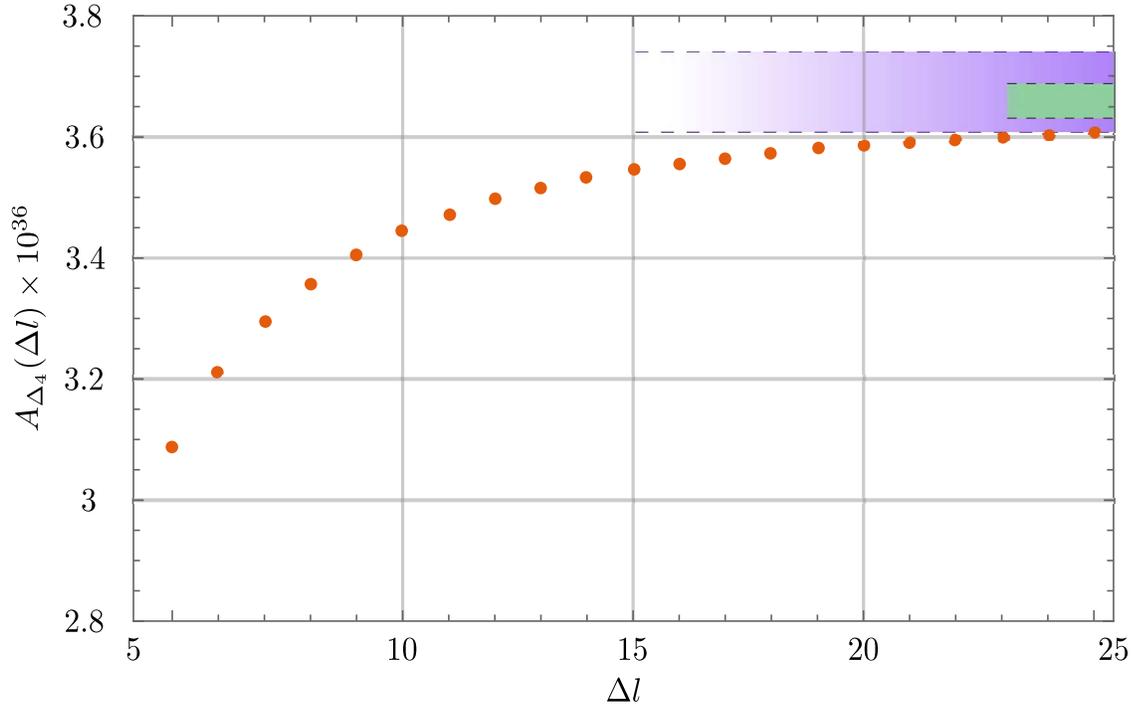}
    \caption{Amplitude $A_{\Delta_4}(\Delta l)$ in function of the cut-off with the band \eqref{eq:bounds-d4} highlighted in purple and the band \eqref{eq:bounds-d4-new} highlighted in green.}
    \label{fig:amplitudedl25}
\end{figure}

The calculation we proposed is not limited to our choice of boundary data.  Using the same technique and adapting the code we compute also the value of the $A_{\Delta_4}$ amplitude for boundary intertwiners $i_b=1,0$ and for other values of the Immirzi parameter $\gamma =1, 0.1$. We summarize the results in the Table~\ref{tab:otheramplitude}

\begin{table}[H]
   \def\arraystretch{1.5}
   \centering
   \begin{tabular}{r|c |c c}
        & $A_{\Delta_4}(25)$ & \multicolumn{2}{c}{$A_{\Delta_4}(25)$}\\
        \hline
        $\gamma=1.2$,$i_b=1$ & $5.06 \cdot 10^{-37}$ & $(5.09 \cdot 10^{-37}$, & $5.11\cdot 10^{-37})$\\
        $\gamma=1.2$,$i_b=0$ & $1.90 \cdot 10^{-37}$ & $(1.92 \cdot 10^{-37}$, & $2.17\cdot 10^{-37})$\\
        \hline
        $\gamma=1.0$,$i_b=2$ & $9.44 \cdot 10^{-34}$ & $(9.51 \cdot 10^{-34}$, & $9.90\cdot 10^{-34})$\\
        $\gamma=1.0$,$i_b=1$ & $1.49 \cdot 10^{-34}$ & $(1.50 \cdot 10^{-34}$, & $1.53\cdot 10^{-34})$\\
        $\gamma=1.0$,$i_b=0$ & $4.95 \cdot 10^{-35}$ & $(5.01 \cdot 10^{-35}$, & $8.50\cdot 10^{-35})$\\
        \hline
        $\gamma=0.1$,$i_b=2$ & $4.28 \cdot 10^{-24}$ & $(4.33 \cdot 10^{-24}$, & $5.49\cdot 10^{-24})$\\
        $\gamma=0.1$,$i_b=1$ & $1.24 \cdot 10^{-24}$ & $(1.26 \cdot 10^{-24}$, & $1.64\cdot 10^{-24})$\\
        $\gamma=0.1$,$i_b=0$ & $2.33 \cdot 10^{-25}$ & $(2.38 \cdot 10^{-25}$, & $2.79\cdot 10^{-25})$\\
        \hline
   \end{tabular}
   \caption{Numerical calculation of the amplitude  $A_{\Delta_4}$ with different boundary data. The boundary spins are all equal $j=1$ and the cut-off $\Delta l =25$.}
   \label{tab:otheramplitude}
\end{table}


\section{Acknowledgments}

This work was made possible through the support of the  FQXi  Grant  FQXi-RFP-1818 and of the ID\# 61466 grant from the John Templeton Foundation, as part of the ``The Quantum Information Structure of Spacetime (QISS)'' Project (\href{qiss.fr}{qiss.fr}). This work was also supported by the Natural Science and Engineering Council of Canada (NSERC) through the Discovery Grant "Loop Quantum Gravity: from Computation to Phenomenology". We also acknowledge the Shared Hierarchical Academic Research Computing Network (SHARCNET) and Compute Canada (\href{https://www.computecanada.ca/}{www.computecanada.ca}) for granting access to their high-performance computing resources. 

We thank F. Gozzini for very insightful comments on the numeric section of our first draft.

We acknowledge the Anishinaabek, Haudenosaunee, L\=unaap\'eewak and Attawandaron peoples, on whose traditional lands Western University is located.

\begin{appendices}



\section{\SUT\ toolbox}
\label{app:SU2}
The group \SUT{} is the group of $2\times 2$ complex matrices with unit determinant that satisfy the unitarity condition
\begin{equation}
\det(u)=1 \ , \ \text{ and } \ \ u^{-1} = u^{\dagger} \ , \ \forall u\in SU(2) \ .
\end{equation}
The group is homomorphic to the rotation group $SO(3)$ and is generated by the angular momentum algebra $L_i$ with $i=1,2,3$ satisfying the commutation relations
\begin{equation}
\label{eq:sutalgebra}
\left[L_{i},L_{j}\right]=i\epsilon_{ijk}L_{k} \ .
\end{equation}
In the fundamental representation $L_i=\sigma_i/2$ where $\sigma_i$ are the standard Pauli matrices. 
The Casimir operator is $L^2 = \vec{L}\cdot\vec{L}$ and the unitary irreducible representations are labeled by a spin $j\in \mathbb{N}/2$ a half-integer and are $2j +1$ dimensional. The canonical basis for these representations diagonalizes the operator $L_3$
\begin{equation}
  L^2\ket{j,m}= j(j+1)\ket{j,m} \ , \qquad L_3\ket{j,m}= m\ket{j,m} \ .
\end{equation}
In this basis the matrix elements of the group are given by the Wigner matrices
\begin{equation}
\label{eq:Wigner_matrix_u}
D^j_{mn}(u) \equiv \bra{j,m} u \ket{j,n} \ .
\end{equation}
Their explicit expression and properties can be found in \cite{book:varshalovic} and we will not report them. 

\medskip

In this work, we compute integrals of products of \SUT{} representation matrices in terms of \SUT{} invariants. We will introduce the minimal amount of tools needed and the graphical method to perform the calculations. We do not want to provide a complete introduction to recoupling theory and its graphical method that are worth books and reviews on their own \cite{Makinen:2019rou,Pierre_notes_LQG,GraphMethods}. 
We use a graphical notation that is completely analogous to the one introduced in Section~\ref{sec:amplitude}.

We associate an oriented line to each \SUT{} representation matrix. We decorate the line with a spin label and a box containing the group element
\begin{align}
\label{eq:irrepsut_graphical}
 D^{j}_{m n}(u)  \hspace{1em} & = \hspace{1em} \raisebox{-0.45\height}{\includegraphics[scale=1]{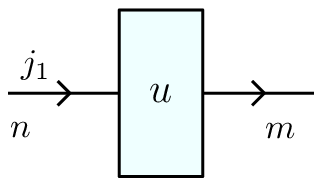}} \ .    
\end{align}
We contract two representations summing over the magnetic indices by connecting the two lines. We compute the integral over \SUT{} using the unique invariant measure over the group (the Haar measure $\dd u$ \cite{Pierre_notes_LQG}). The explicit form of the measure depends on the parametrization used for the group. We collect the boxes corresponding to the same group elements. In the following, we will always imply the integration over all the group elements in the boxes.

\medskip

The integral of the product of two representation matrices is given by 
\begin{align}
  \label{eq:twowigner}
  \int \dd u D^{j_1}_{m_1 n_1}(u)D^{j_2}_{m_2 n_2}(u) = \frac{1}{2j_1 +1} \delta_{j_1j_2} (-1)^{2j_1 - m_1-n_1} \delta_{-m_1 m_2}\delta_{-n_1 n_2} = \frac{1}{2j_1 +1} \delta_{j_1j_2} \epsilon^{j_1}_{m_1m_2} \epsilon^{j_1}_{n_1n_2} \ ,
\end{align}
where we defined the tensor $\epsilon^{j_1}_{m_1m_2}\equiv(-1)^{j_1-m_1} \delta_{-m_1 m_2}$, the unique invariant tensor in the product of two $j_1$ representations. The $\epsilon$ tensor squares to 
\begin{equation}
\label{squareepsilon}
\sum_{m_2} \epsilon^{j_1}_{m_1m_2} \epsilon^{j_1}_{m_2m_3} = 
\sum_{m_2} (-1)^{j_1-m_1} \delta_{-m_1 m_2} (-1)^{j_1-m_2} \delta_{-m_2 m_3} = 
(-1)^{2j_1-m_1 + m_3} \delta_{m_1 m_3} = (-1)^{2j_1} \delta_{m_1 m_3} \ ,
\end{equation}
and has the symmetry property $\epsilon^{j_1}_{m_1m_2} = (-1)^{2j_1} \epsilon^{j_1}_{m_2m_1}$. We use the graphical representation to write \eqref{eq:twowigner} as
\begin{align}
  \label{eq:twowigner-graphical}
  \raisebox{-0.45\height}{\includegraphics[scale=1]{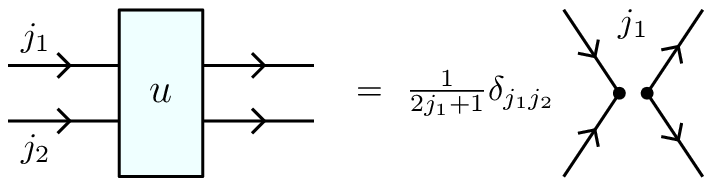}} \ .    
\end{align}
The invariance property of the tensor $\epsilon^{j_1}_{m_1m_2}$ means
\begin{equation}
  \label{eq:2invariant}
  \sum_{n_1, n_2} D^{j_1}_{m_1 n_1}(u)D^{j_1}_{m_2 n_2}(u) \epsilon^{j_1}_{n_1n_2}=\epsilon^{j_1}_{m_1m_2} \ .
\end{equation}
From \eqref{eq:2invariant} we can derive the property of Wigner matrices
\begin{equation}
  \label{eq:wignerinverse}
  \sum_{m_2 n_2}\epsilon^{j_1}_{n_1 m_2} \epsilon^{j_1}_{m_1 n_1} D^{j_1}_{m_2 n_2}(u)   =  D^{j_1}_{m_1 n_1}(u^{-1}) \ .
\end{equation}
Using this property we can also perform integrals where an inverse group element appears
\begin{align}
  \label{eq:twowignerinverse}
  \int \dd u D^{j_1}_{n_1 m_1}(u^{-1})D^{j_2}_{m_2 n_2}(u) =& \raisebox{-0.45\height}{\includegraphics[scale=1]{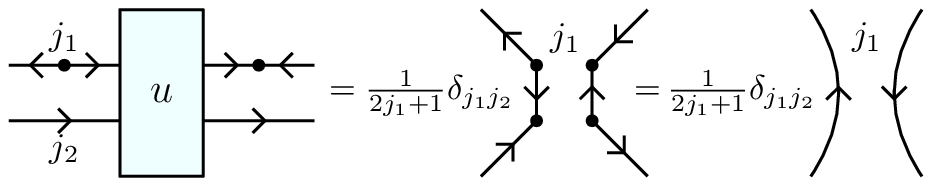}} \\ &= \frac{1}{2j_1+1}  \delta_{j_1j_2}\delta_{m_1 m_2}\delta_{n_1 n_2} \ .    
\end{align}
For simplicity, we will merge the $\epsilon$ tensors with the box in the following. At first sight, it could appear as an ambiguity since one line will have a group element $u$ in the box, while the line with the opposite orientation $u^{-1}$ and there is no indication of which is which. However, we are integrating over $u$. Therefore, the name we give the group element is irrelevant. The important information is contained in the relative polarity: one group element is the inverse of the other.

Using this convention and the square property \eqref{squareepsilon}, in any closed diagram, inverting the orientation of a line (without group elements) results into a phase $(-1)^{2j}$.
\begin{align}
  \label{eq:lineinversion}
  \raisebox{-0.45\height}{\includegraphics[scale=1]{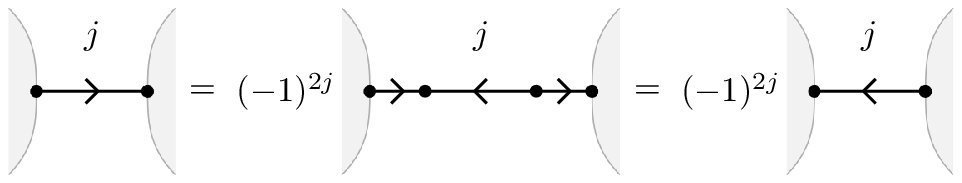}} \ .
\end{align}

\medskip

The integral of the product of three representation matrices is given by 
\begin{align}
  \label{eq:threewigner}
  \int \dd u D^{j_1}_{m_1 n_1}(u)D^{j_2}_{m_2 n_2}(u)D^{j_3}_{m_3 n_3}(u) = \Wthree{j_1}{j_2}{j_3}{m_1}{m_2}{m_3} \Wthree{j_1}{j_2}{j_3}{n_1}{n_2}{n_3} \ .
\end{align}
The tensors appearing in \eqref{eq:threewigner} are the Wigner $(3jm)$ symbols, the unique invariant tensor (or three valent \emph{intertwiner}) in the tensor product of three \SUT{} representations.
\begin{equation}
  \label{eq:threewignergraphical}
  \sum_{n_1, n_2,n_2} D^{j_1}_{m_1 n_1}(u)D^{j_2}_{m_2 n_2}(u)D^{j_3}_{m_3 n_3}(u) \Wthree{j_1}{j_2}{j_3}{n_1}{n_2}{n_3}=\Wthree{j_1}{j_2}{j_3}{m_1}{m_2}{m_3} \ .
\end{equation}
The $(3jm)$ has the following symmetry properties (see \cite{Pierre_notes_LQG,GraphMethods,book:varshalovic} for an exhaustive list)
\begin{align}
  \Wthree{j_1}{j_2}{j_3}{n_1}{n_2}{n_3} = \Wthree{j_2}{j_3}{j_1}{n_2}{n_3}{n_1} =  (-1)^{j_1+ j_2 + j_3} \Wthree{j_1}{j_3}{j_2}{n_1}{n_3}{n_2} \ ,
\end{align}
and vanishes unless the selection rules are satisfied
\begin{align}
\label{eq:selection_rules_3jm_symbol}
m_1 + m_2 + m_3 & = 0 \ , & |j_1 - j_2|  \leq  j_3  \leq  & j_1 + j_2 \ , & j_1 + j_2 + j_3 \in \ & \mathbb{N} \ .
\end{align}
In the graphical representation \eqref{eq:threewigner} is
\begin{align}
  \raisebox{-0.45\height}{\includegraphics[scale=1]{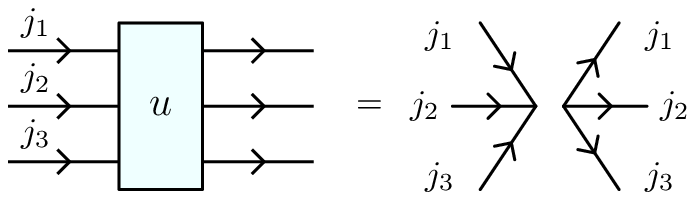}} \ ,    
\end{align}
where for the $(3jm)$ symbol
\begin{equation}
  \Wthree{j_1}{j_2}{j_3}{m_1}{m_2}{m_3} = \raisebox{-0.45\height}{\includegraphics[scale=1]{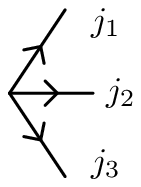}}  \ ,
\end{equation}
we read the spins in clockwise order if all the arrows are outgoing and in anti-clockwise order if all the arrows are ingoing. In the standard \SUT{} graphical calculus, this is usually indicated with a sign next to the node \cite{GraphMethods,book:varshalovic,Makinen:2019rou}. For our calculations, this is unnecessary, and we avoid adding this extra layer of complexity. Similarly to \eqref{eq:twowignerinverse} we have
\begin{align}
  \label{eq:threewignerinverse}
  \int \dd u D^{j_1}_{n_1 m_1}(u^{-1})D^{j_2}_{m_2 n_2}(u)D^{j_3}_{m_3 n_3}(u) =& \raisebox{-0.35\height}{\includegraphics[scale=1]{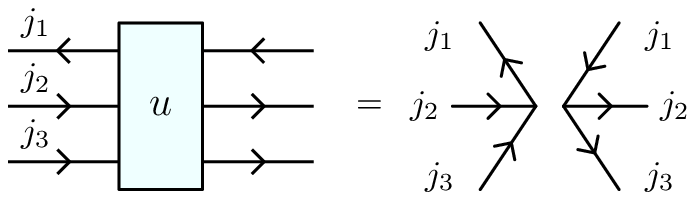}} =\\
  & (-1)^{j_1-n_1} \Wthree{j_1}{j_2}{j_3}{-n_1}{n_2}{n_3}  (-1)^{j_1-m_1}\Wthree{j_1}{j_2}{j_3}{-m_1}{m_2}{m_3} \ . \nonumber
\end{align}
The $(3jm)$ satisfy the orthogonality relation  
\begin{align}
  \label{eq:3jorthogonality}
  \sum_{m_2,m_3}\Wthree{j_1}{j_2}{j_3}{m_1}{m_2}{m_3}\Wthree{j_1'}{j_2}{j_3}{m_1'}{m_2}{m_3} = \raisebox{-0.35\height}{\includegraphics[scale=1]{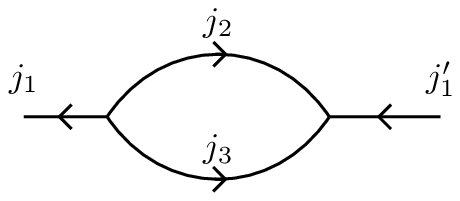}}  = \frac{1}{2j_1+1}\delta_{j_1 j_1'} \delta_{m_1 m_1'} \ ,
\end{align}
and are normalized to 1 
\begin{align}
  \label{eq:3jnormalization}
  \sum_{m_1,m_2,m_3}\Wthree{j_1}{j_2}{j_3}{m_1}{m_2}{m_3}\Wthree{j_1}{j_2}{j_3}{m_1}{m_2}{m_3} = \raisebox{-0.35\height}{\includegraphics[scale=1]{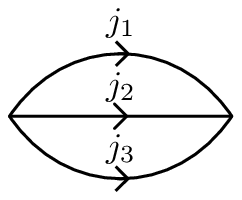}}  = 1 \ .
\end{align}

\medskip 

We also compute the integral of the product of four representation matrices with the tools we provided using a trick 
\begin{align}
  \label{eq:fourwigner}
  \int \dd u D^{j_1}_{m_1 n_1}(u)D^{j_2}_{m_2 n_2}(u)D^{j_3}_{m_3 n_3}(u) D^{j_4}_{m_4 n_4}(u) = \int \dd u\dd v D^{j_1}_{m_1 n_1}(u)D^{j_2}_{m_2 n_2}(u) \delta(uv^{-1}) D^{j_3}_{m_3 n_3}(v) D^{j_4}_{m_4 n_4}(v) \ ,
\end{align}
where we doubled the number of integrals inserting a delta function. The delta function can be expanded as a sum of Wigner functions \cite{Rovelli2015}
\begin{equation}
\delta(uv^{-1})  = \sum_{i} (2j+1) \sum_m D_{mm}^i (uv^{-1})= \sum_{i} (2i+1) \sum_{m,n} D_{mn}^i (u)D^i_{nm}(v^{-1}) \ .
\end{equation}
Using the graphical representation and the properties \eqref{eq:threewignergraphical} and \eqref{eq:threewignerinverse} we compute \eqref{eq:fourwigner}
\begin{align}
\label{eq:fourwignerintegral}
  \raisebox{-0.45\height}{\includegraphics[scale=1]{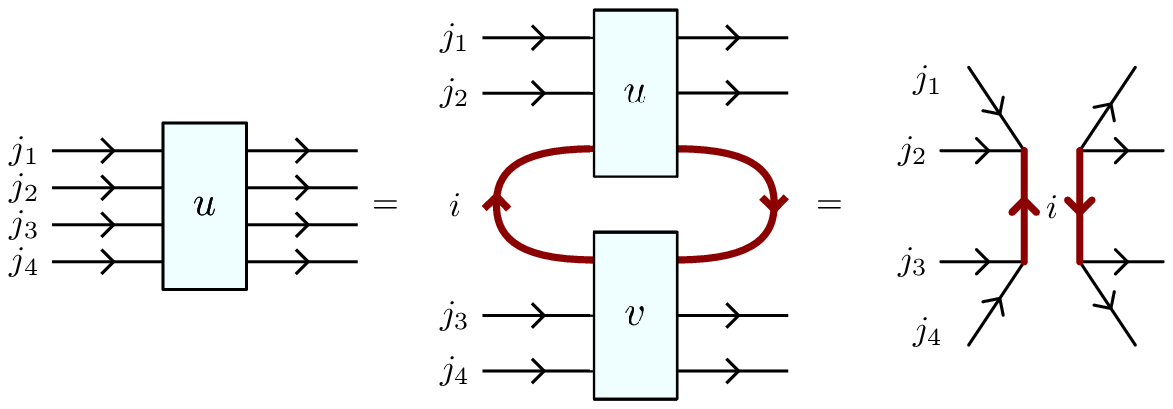}} \ ,
\end{align}
where we used a red thick line to imply a summation weighted by the dimensional factor $(2i+1)$ and we define the $(4jm)$ symbols (invariant tensor or four valent intertwiner) as
\begin{equation}
  \label{eq:4jm}
  \raisebox{-0.45\height}{\includegraphics[scale=1]{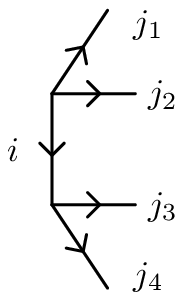}} = \Wfour{j_1}{j_2}{j_3}{j_4}{m_1}{m_2}{m_3}{m_4}{i} = \sum_m (-1)^{i-m} \Wthree{j_1}{j_2}{i}{m_1}{m_2}{m} \Wthree{i}{j_3}{j_4}{-m}{m_3}{m_4} \ .
\end{equation}
The way we grouped representations together in \eqref{eq:fourwigner} is completely arbitrary. The definition \eqref{eq:4jm} corresponds to the choice of coupling (also called recoupling basis) of the representation of spins $j_1$ and $j_2$ (ore equivalently spins $j_3$ and $j_4$).
The orthogonality condition \eqref{eq:3jorthogonality} of the $(3jm)$ symbols imply the normalization of the $(4jm)$ symbols 
\begin{align}
  \label{eq:4jmtheta}
  \raisebox{-0.45\height}{\includegraphics[scale=1]{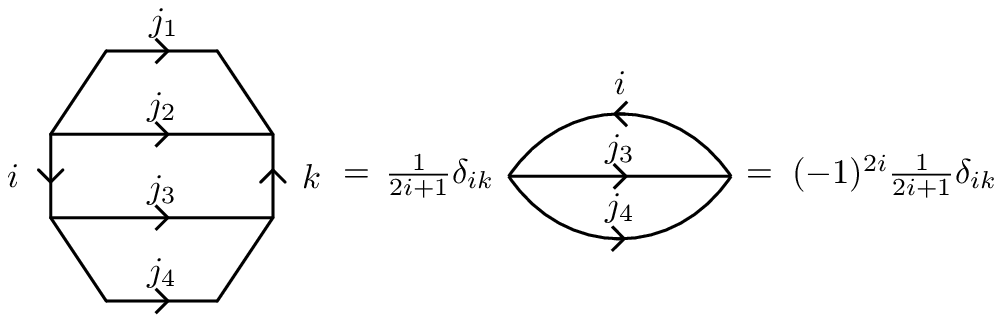}} \ .
\end{align}

\medskip

The contraction of two $(4jm)$ symbols in different recoupling basis forms another notable \SUT{} invariant called the $\{6j\}$ symbol.
\begin{align}
  \raisebox{-0.45\height}{\includegraphics[scale=1]{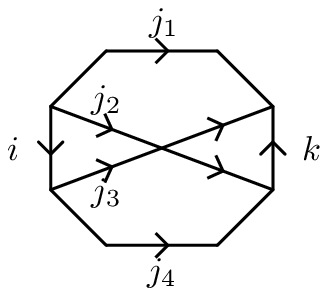}} = (-1)^{j_2+j_3 +i+k}  \Wsix{j_1}{j_2}{i}{j_4}{j_3}{k} \ .
\end{align}
The $\{6j\}$ symbol in terms of $(3jm)$ symbols can be written in a canonical from as
\begin{align}
\label{eq:6j_symbol}
  \Wsix{j_1}{j_2}{j_3}{j_4}{j_5}{j_6} 
  & = \sum\limits_{m_1 \dots m_6} (-1)^{\sum\limits_{i=1}^{6} ( j_i - m_i ) } \Wthree{j_1}{j_2}{j_3}{m_1}{m_2}{-m_3} \Wthree{j_1}{j_5}{j_6}{-m_1}{m_5}{m_6} \times \nonumber \\
  & \hspace{6mm} \times \Wthree{j_4}{j_5}{j_3}{m_4}{-m_5}{m_3} \Wthree{j_4}{j_2}{j_6}{-m_4}{-m_2}{-m_6} \ .
\end{align}
For a numerical evaluation, it is not convenient to write the $\{6j\}$ symbol as in \eqref{eq:6j_symbol}. It is much more efficient to rely on libraries that compute and store Wigner $\{6j\}$ symbols optimally using recursion and symmetry properties, such as \texttt{wigxjpf} and \texttt{fastwixj} \cite{Wigxjpf_library, fastwigxj_related}. 

\medskip

Another higher-order invariant that appears in our calculations is the irreducible $\{15j\}$ symbol of the first kind (following the classification of \cite{GraphMethods}). We can write it both graphically and in terms of $\{6j\}$ symbols as:
\begin{align}
  \label{eq:15jsymbol}
  \left \{ \begin{array}{ccccc} j_1 & j_2 & j_3 & j_4 & j_5 \nonumber \\  
  l_1 & l_2 & l_3 & l_4 & l_5 \nonumber \\ 
  k_1 & k_2 & k_3 & k_4 & k_5 \end{array}\right \}   & = \raisebox{-0.45\height}{\includegraphics[scale=1]{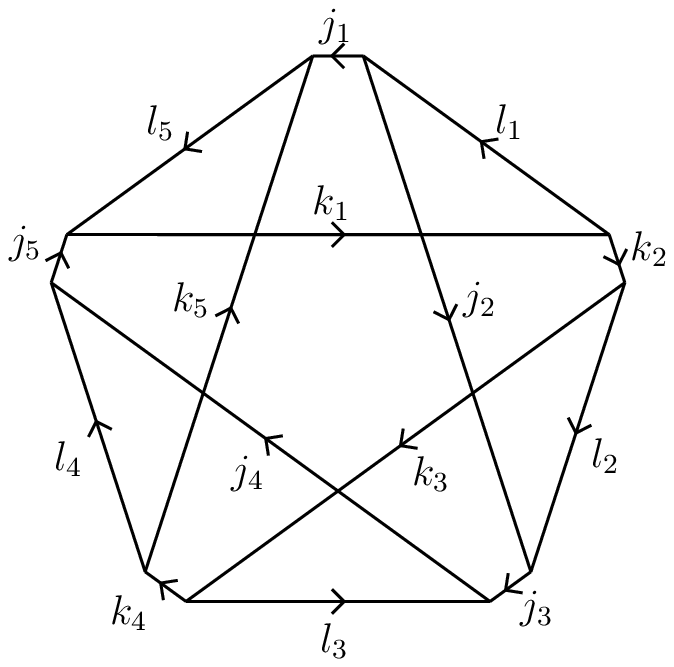}} \\ 
  &=  \ \ (-1)^{\sum_{i=1}^5 j_i + l_i +k_i} \sum_x (2 x +1) \Wsix{j_1}{k_1}{x}{k_2}{j_2}{l_1} \Wsix{j_2}{k_2}{x}{k_3}{j_3}{l_2} \nonumber \\ & \hspace{6mm }\times \Wsix{j_3}{k_3}{x}{k_4}{j_4}{l_3} \Wsix{j_4}{k_4}{x}{k_5}{j_5}{l_4} \Wsix{j_5}{k_5}{x}{j_1}{k_1}{l_5}  \ .
\end{align}
It must be emphasized that the $\{15j\}$ symbol \eqref{eq:15jsymbol} is not the most convenient choice from a numerical point of view. In fact, it is possible to choose the recoupling scheme in order to obtain reducible $\{15j\}$ symbols (see \cite{article:Dona_etal_2018_SU2_graph_invariants} for an example), whose evaluation is much faster. However, since this aspect is not the most critical part of the performance, we prefer to have a pleasantly symmetrical symbol and sacrifice some efficiency. This also simplifies computations of spin foams transition amplitudes with many vertices, since the basis choice in the recoupling on one edge affects both vertices it connects. Therefore, choosing a symmetric $\{15j\}$ symbol as in \eqref{eq:15jsymbol}, we are sure that the recoupling is consistent in every vertex.

%

\section{\sltc\ toolbox}
\label{app:sltc}
The group \sltc{} is the group of $2\times 2$ complex matrices with unit determinant. The group is homomorphic to the proper Lorentz group (the Lorentz group part that preserve the sign of the time component) \cite{CarmeliBook, Ruhl:1970fk}. 

\smallskip

The algebra of \sltc{} is generated by spatial rotations and boosts $L_i$ and $K_i$ satisfying the commutation relations
\begin{equation}
\label{eq:sl2calgebra}
\left[L_{i},L_{j}\right]=i\epsilon_{ijk}L_{k} \ ,\qquad
\left[L_{i},K_{j}\right]=i\epsilon_{ijk}K_{k} \ ,\qquad
\left[K_{i},K_{j}\right]=-i\epsilon_{ijk}L_{k} \ .
\end{equation}
In the spinorial representation $L_i=\sigma_i/2$ and $K_i = i \sigma_i/2$ where $\sigma_i$ are the standard Pauli matrices. 
The two Casimir operators are $K^2 - L^2$ and $\vec{K}\cdot\vec{L}$. The unitary irreducible representations in the principal series are labeled by $\rho$ a real number and $k$ a half-integer. In these representations the Casimirs assume the values 
\begin{equation}
\label{eq:CasimirSL2C}
\left(K^2 - L^2\right)\ket{\rho,k}= (\rho^2-k^2+1)\ket{\rho,k} \ , \qquad \vec{K}\cdot\vec{L}\ket{\rho,k}=\rho k \ket{\rho,k} \ .
\end{equation}

The generic unitary representation $(\rho,k)$ is infinite dimensional since the group is non-compact. However, we can decompose the representation $(\rho,k)$ in an infinite number of \SUT{} representations that diagonalize $L^2$ with different values of the spin $j$
\begin{equation}
\label{eq:SU2TowerSL2C}
(\rho,k) = \bigoplus_{j\geq k} {j} \ .
\end{equation}
The definition of the EPRL model is based on the canonical basis of $(\rho,k)$. In this basis we diagonalize $L^2$ and $L_3$ 
\begin{equation}
L^2\ket{\rho,k;j,m}= j(j+1)\ket{\rho,k;j,m} \ , \qquad L_3\ket{\rho,k;j,m}= m\ket{\rho,k;j,m} \ .
\end{equation}
with $j\geq k$ and $m = -j,\ldots,j$ . 

\smallskip
The Cartan parametrization  \cite{Ruhl:1970fk, Speziale2016} of the group \sltc{} is given by the map
\begin{equation}
\begin{split}
\label{eq:CartanSL2C}
  g = ue^{\frac{r}{2} \sigma_3} v^{-1}
   \ ,
\end{split}
\end{equation}
where $u,v\in SU(2)$, $r\in[0,\infty)$ is the rapidity and $\sigma_3$ is the diagonal Pauli matrix the generator of boosts along the $z$ axis. The Haar measure with respect to this parametrization is \cite{Ruhl:1970fk, Speziale2016} 
\begin{equation}
\label{eq:HaarSL2C}
 \dd g =  \frac{1}{4\pi} \sinh^2 r \ \dd r \ \dd u\  \dd v \ .
\end{equation}
Using the Cartan parametrization \eqref{eq:CartanSL2C} the matrix elements of a group element $g$ in the canonical basis reads
\begin{equation}
D_{jmln}^{\rho,k}(g) \equiv \bra{\rho,k;j,m} g \ket{\rho,k;l,n} =  D_{jmln}^{\rho,k}(ue^{\frac{r}{2} \sigma_3}v^{-1}) = \sum_{a,a'} D_{ma}^{j}(u)D_{jala'}^{\rho,k}(e^{\frac{r}{2} \sigma_3}) D_{a'n}^{l}(v^{-1}) \ .
\end{equation}
The subgroup $SU(2)\subset SL(2,\mathbb{C})$ is generated by $\vec{L}$ and its matrix elements are given by \SUT{} Wigner matrices \eqref{eq:Wigner_matrix_u}
\begin{equation}
\label{eq:sltcsu2}
D_{jmln}^{\rho, k}(u)=\bra{\rho,k;j,m} u \ket{\rho,k;l,n} = \delta_{jl} D^j_{mn}(u) \ \text{ where } \ u \in SU(2) \  .
\end{equation}
Moreover, $e^{\frac{r}{2} \sigma_3}$ is diagonal, therefore $D_{jmln}^{\rho,k}(e^{r\sigma_3})= \delta_{aa'} D_{jala}^{\rho,k}(e^{r\sigma_3}) \equiv \delta_{aa'} d_{jla}^{\rho,k}(r)$ where $d_{jla}^{\rho,k}$ are called reduced matrix elements of \sltc. Summarizing
\begin{equation}
\label{eq:CartanDSL2C}
D_{jmln}^{\rho,k}(g) = \sum_{a} D_{ma}^{j}(u)\ d_{jla}^{\rho,k}(r)\ D_{an}^{l}(v^{-1}) \ .
\end{equation}
The expression for $d_{jlm}^{\rho,k}(r)$ was given in \cite{Dao:1967ri,Rashid:1979xv,Basu:1977ii, Ruhl:1970fk, Speziale2016}
\begin{equation}\label{eq:dSL2C}
\begin{split}
d_{jlm}^{\rho,k}(r)=(-1)^{j-l}&\sqrt{\frac{\left(i\rho-j-1\right)!\left(j+i\rho\right)!}{\left(i\rho-l-1\right)!\left(l+i\rho\right)!}}\frac{\sqrt{(2j+1)(2l+1)}}{(j+l+1)!}e^{(i\rho-k-m-1)r}\\
&\sqrt{(j+k)!(j-k)!(j+m)!(j-m)!(l+k)!(l-k)!(l+m)!(l-m)!}\\
&\sum_{s,t}(-1)^{s+t}e^{-2tr}\frac{(k+s+m+t)!(j+l-k-m-s-t)!}{t!s!(j-k-s)!(j-m-s)!(k+m+s)!(l-k-t)!(l-m-t)!(k+m+t)!}\\
&{}_{2}F_{1}\left[\{l-i\rho+1,k+m+s+t+1\},\{j+l+2\};1-e^{-2r}\right]
\end{split}
\end{equation}
where ${}_{2}F_{1}$ is the Gauss hypergeometric function. The phase used in \eqref{eq:dSL2C} is the same introduced in \cite{Speziale2016}, which ensures the reality of the booster function \eqref{eq:boosterdef}. The reduced matrix elements \eqref{eq:dSL2C} satisfy the following relation:
\begin{equation}
\label{eq:d_sl2c_overline}
\overline{d^{\rho,k}_{j l m}(r)} = (-1)^{j-l} d^{\rho,k}_{j l - m }(r) \ .
\end{equation}
As a consequence, the matrices \eqref{eq:sltcsu2} have the property:
\begin{equation}
\label{eq:sl2c_D_overline}
\overline{D^{\rho,k}_{j m l n}(g)} = (-1)^{j-l+m-n} D^{\rho,k}_{j - m l - n}(g) \ .
\end{equation}
We can write the \sltc{} matrix elements of $g^{-1}$ as:
\begin{equation}
\label{eq:sl2c_equality_chain}
D^{\rho, k}_{l n j m}(g^{-1})= (-1)^{j-l+m-n} D^{\rho, k}_{j - m l - n }(g) = \sum_{a} (-1)^{j-l+m-n}\ D^{j}_{-m a}(u)\ d^{\rho, k}_{j l a}(r)\  D^{l}_{a -n}(v^{-1})  \ ,  
\end{equation}
where in the first equality we used \eqref{eq:sl2c_D_overline} (in addition to the \sltc{} irrep properties) and in the second one \eqref{eq:CartanDSL2C}. Since there are no phases depending on the summed index, we conclude that the orientation of the $(4jm)$ spins 
in the booster function \eqref{eq:boosterdef} is irrelevant. This justifies the fact that we draw the latter without arrows.


\section{Approximation of a convergent series}
\label{app:seriesproof}
\new{In this appendix, we provide further details on the extrapolation scheme used in Section \ref{sec:numerical}. This is analogous to the more general Aitken's delta-squared process \cite{Aitken:1927}, which accelerate the rate of convergence of a sequence providing a good approximation technique}.
Consider the series $S = \sum_n^\infty a_n $ and cut-offed sum $S_N = \sum_n^N a_n$. By definition the series is the limit of $S_N$ for infinite cut-off 
\begin{equation}
  S=\lim_{N\to\infty} S_N = \lim_{N\to\infty}  \sum_n^N a_n  = \sum_n^\infty a_n \ .
\end{equation}
Suppose that the sequence $a_n$ is positive and, from a certain point onwards, increasing such that 
\begin{equation}
  \lim_{N\to\infty}c_N \equiv  \lim_{N\to\infty} \frac{a_{N}}{a_{N-1}} =\lim_{N\to\infty} \frac{S_{N}-S_{N-1}}{S_{N-1}-S_{N-2}} \equiv L <1 \ ,
\end{equation}
where the ratios increase to $L$. The series $S$ is convergent by the ratio test since the ratios are increasing:
\begin{equation}
  c_N = \frac{a_{N}}{a_{N-1}} < \frac{a_{k}}{a_{k-1}} \ , \qquad \forall k > N \ . 
\end{equation}
Hence we have $a_{N+1} = a_{N} \frac{a_{N+1}}{a_{N}} > a_{N} c_N$, $a_{N+2} > a_{N+1} c_N> a_N c_N^2$, and in general $a_{N+m} > a_N c_N^m$ for $m>0$.
We can provide a bound on the series observing that 
\begin{equation}
  S-S_N = \sum_{n=N+1}^\infty a_n  = \sum_{m=1}^\infty a_{N+m} > \sum_{m=1}^\infty a_N c_N^m = a_N \frac{c_N}{1-c_N} \ .
\end{equation}
Similarly, we have that $\frac{a_{k}}{a_{k-1}}<L$ $\forall k > N$ by definition of $L$ and monotonicity of the ratios.
\begin{equation}
  S-S_N = \sum_{n=N+1}^\infty a_n  = \sum_{m=1}^\infty a_{N+m} < \sum_{m=1}^\infty a_N L^m = a_N \frac{L}{1-L} \ .
\end{equation}
Summarizing, we have an estimate from above and below of the value of the series as 
\begin{equation}
  \label{eq:bounds-0}
  S_N + a_N \frac{c_N}{1-c_N} < S < S_N + a_N \frac{L}{1-L} \ .
\end{equation}
If the ratios are decreasing instead of increasing, we obtain an estimate analog to \eqref{eq:bounds-0} but with inequalities operators inverted. Let's focus on  \eqref{eq:bounds-0} since it is the case relevant for the cut-off approximation presented in Section~\ref{sec:numerical}. We rewrite \eqref{eq:bounds-0} in terms of cut-offed sums as
\begin{equation}
  \label{eq:bounds}
 \frac{S_N S_{N-2} - S_{N-1}^2}{S_N - 2S_{N-1}+S_{N-2}} < S <  \frac{S_N-S_{N-1}L}{1-L} \ .
\end{equation}

\medskip

Often, in real-world physical applications, the analytical expression of $a_n$ is very complicated. We cannot compute $S$ but can still calculate the cut-offed sums $S_N$ with $N$ as large as our numerical computational resources allow. What is the best approximation of $S$ we can find? We will assume that we know $S$ is convergent (so that the question is well-posed) and that the ratios $c_N$ increase. We want to use the inequalities \eqref{eq:bounds}. We can numerically compute the left-hand side of the inequality. What about the right-hand side? 
The convergence of $S$ ensure that $\displaystyle \lim_{N\to\infty}c_N =L$ exists, however in general we cannot compute the value of $L$. At the moment, there is no clear strategy on how to compute $L$. In this work, we will consider two possibilities. None of them is optimal, and we leave improvements to future work. Notice that no matter what approximation we decide to adopt to compute $L$ the right inequality of \eqref{eq:bounds} will not hold anymore. 

\medskip

For example, we can approximate $L$ with the largest available ratio $L\approx c_N$. In particular, if we insist and substitute in \eqref{eq:bounds} the approximation $L\approx c_N$, the right quantity becomes equal to the left one. We have to content ourselves with a lower bound estimate of the amplitude given 
\begin{equation}
  \label{eq:bounds-1}
 S\gtrapprox\frac{S_N S_{N-2} - S_{N-1}^2}{S_N - 2S_{N-1}+S_{N-2}}  \ .
\end{equation}
The estimate \eqref{eq:bounds-1} is very similar to the strategy used in \cite{frisoni2021studying,frisoni2021numerical}. We clarified that it is a lower bound. Another possibility is to use the sequence of ratios $c_N$ computed numerically to estimate the value of $L$. This extrapolation is slightly dangerous since its accuracy depends on how large we can take the cut-off $N$. Of course, this is in addition to the lower bound \eqref{eq:bounds-1}. 

\medskip 

In the following, we provide a concrete toy model example. Consider the series
\begin{equation}
  S= \sum_{n=1}^\infty \frac{1}{n+1} (9/10)^n = \frac{10}{9}\log(10) -1 \approx 1.558 \ .
\end{equation}
This series is exactly summable in terms of the log function. Nevertheless, we want to approximate the series pretending not to know how to sum it, ignoring the fact that any analytical calculation is straightforward, an relying only on numerical tools. Let's assume that the largest possible cut-off we have access to is $N=15$. We can compute the cut-offed sums 
\begin{align}
  S_{15} &\approx 1.480 \ , &S_{14} &\approx 1.467 \ , &S_{13} &\approx 1.452 \ .
\end{align} 
We can immediately apply \eqref{eq:bounds-1} to obtain
\begin{equation}
  \label{eq:examplebound}
 S\gtrapprox\frac{S_{15} S_{13} - S_{14}^2}{S_{15} - 2S_{14}+S_{13}}  \approx 1.549\ .
\end{equation}
The largest cut-offed sum is $5\%$ off the real value while the lower bound approximation \eqref{eq:examplebound} is closer, being only $0.6\%$ off. The numerical values for the ratios are

\medskip 
\noindent
\begin{tabular}{llllllllllllll}
  N    & $3$ & $4$ & $5$ & $6$ & $7$ & $8$ & $9$ & $10$ & $11$ & $12$ & $13$ & $14$ & $15$\\
  $c_N$ & $0.675$ & $0.720$ & $0.750$ & $0.771$ & $0.788$ & $0.800$ & $0.810$ & $0.818$ & $0.825$ & $0.831$ & $0.836$ & $0.840$ & $0.844$
\end{tabular}

\medskip 

We extrapolate the limit at infinity of the ratios $L$ fitting the data using the first few terms of an inverse power law and keeping the constant term. It is a cheap and dirty way of extrapolating, and one should be more careful. However, it is more than enough for our purposes. We use Wolfram's Mathematica built-in \texttt{Fit} method to perform the fit and find $L\approx 0.889$. If we substitute it in \eqref{eq:bounds}, keeping in mind the approximations we are making, we find the estimate
\begin{equation}
  S\lessapprox\frac{S_{15}-S_{14}L}{1-L}\approx 1.583 \ ,
\end{equation}
which is $1.6\%$ larger than the actual value. Combining the two estimates, we obtain a range for the series $S\in [1.549,1.583]$.

\end{appendices}


\end{document}